\documentclass[10pt]{article}

\usepackage{amsmath}
\usepackage{amssymb}

\usepackage{graphicx}

\usepackage{cite}

\usepackage{color} 
\usepackage{psfrag}

\usepackage[labelfont=bf,labelsep=period,justification=raggedright]{caption}

\makeatletter
\renewcommand{\@biblabel}[1]{\quad#1.}
\makeatother

\date{}

\begin{document}

\begin{flushleft}
{\Large
\textbf{Genetic Classification of Populations  using Supervised Learning}
}
\\
\bigskip
Michael Bridges$^{1}$, 
Elizabeth A. Heron$^{2}$, 
Colm O'Dushlaine$^{2}$, 
Ricardo Segurado$^{2}$, 
The International Schizophrenia Consortium (ISC)$^{\dagger}$,
Derek Morris$^{2}$, 
Aiden Corvin$^{2}$, 
Michael Gill$^{2}$, 
Carlos Pinto$^{2,\ast}$ 
\\
\bigskip
{\bf 1} Astrophysics Group, Cavendish Laboratory, Cambridge, CB3 0HE, UK
\\
{\bf 2} Neuropsychiatric Genetics Research Group, Department of Psychiatry,  Trinity College, Dublin, Ireland
\\
\bigskip
{\bf $\dagger$} The full list of contributors in this consortium is presented in the Supporting Information.
\\
\bigskip
{\bf $\ast$ E-mail: capinto@tcd.ie}
\end{flushleft}

\section*{Abstract}

There are many instances in genetics in which we wish to determine whether two candidate populations are distinguishable on the basis of their genetic
structure. Examples include populations which are geographically separated, case--control studies and quality control (when participants in a study
have been genotyped at different laboratories). This latter application is of particular importance in the era of large scale genome wide association
studies, when collections of individuals genotyped at different locations are being merged to provide increased power. The traditional method for
detecting structure within a population is some form of exploratory technique such as principal components analysis.  Such methods, which do not
utilise our prior knowledge of the membership of the candidate populations. are termed \emph{unsupervised}. Supervised methods, on the other hand are
able to utilise this prior knowledge when it is available.

In this paper we demonstrate that in such cases modern supervised approaches are a more appropriate tool for detecting genetic differences between
populations. We apply two such methods, (neural networks and support vector machines) to the classification of three populations (two from Scotland
and one from Bulgaria). The sensitivity exhibited by both these methods is considerably higher than that attained by principal components analysis and
in fact comfortably exceeds a recently conjectured theoretical limit on the sensitivity of unsupervised methods. In particular, our methods can
distinguish between the two Scottish populations, where principal components analysis cannot. We suggest, on the basis of our results that a
supervised learning approach should be the method of choice when classifying individuals into pre-defined populations, particularly in quality control
for large scale genome wide association studies.

\section*{Introduction}
\label{sec:intro}
The advent of the new large-scale genotyping and sequencing technologies has resulted in unprecedented quantities of data becoming available to the
genetics community. Geneticists are now confronted with new and challenging problems in data analysis and interpretation, and novel approaches and
techniques will be required to fully exploit these new resources.  In view of the fact that other scientific fields have already gone through a
similar process of development, it is likely that cross-disciplinary collaborations in data analysis will yield fruitful results in genetics. This
paper represents such a collaboration.

We apply machine learning techniques previously used in cosmology to the problem of genetic classification. Such techniques involve the use of
automated algorithms to mimic the learning capabilities of animal brains. They have proved extremely useful in the analysis of complex data in
many scientific disciplines. There are two basic approaches -- \emph{supervised} learning, where the data is pre-classified according to some
hypothesis and \emph{unsupervised} learning where the data is unclassified (usually, but not always, because the potential classes are \emph{a priori}
unknown). Genetics has, to date, relied mainly on unsupervised methods, such as principal components analysis (PCA), to classify individuals on the
basis of their genetic data.  

PCA is a standard tool in population genetics, and has been used, for example in a study of 23 European populations \cite{Lao} and more recently of 25
Indian populations \cite {Reich1}.  It is also commonly used in quality control in genetic studies. For example, a dataset destined for a disease
association study may be pre-screened using PCA in order to detect and remove population structure so as to minimise noise in the final study. In many
of the large scale collaborations now being undertaken it is of interest to determine whether genetic differences exist between groups of controls
ascertained from different geographic locations, or genotyped at different laboratories. If the differences are sufficiently small, these groups can
be merged to achieve greater power.  The aim of this work is to demonstrate and quanmtify the superiority of supervised learning techniques when
applied to this problem.

We have adapted two supervised learning algorithms, artificial neural networks (ANN) and support vector machines (SVM) for this purpose. We use sets
of control samples genotyped by the International Schizophrenia Consortium (ISC) \cite{ISC} as our test data. For comparison we also conduct a
conventional PCA analysis.

The paper is organised as follows. In the Methods section we briefly discuss the PCA methodology that we use and give a short introduction to ANNs and
SVMs. We also include a description of the data used for the analysis. The first part of the Results section presents the PCA analysis and
results. The second and third sections describe the ANN and SVM analyses respectively. Finally, the Discussion section contains our interpretation of
the analyses and some suggestions for potential applications of the methods.

\section*{Methods}
\label{sec:methods}

We examine three approaches to the problem of genetic classification, given pre--existing candidate populations.  More precisely, we wish to determine
the confidence with which the individuals in these populations can be distinguished on the basis of their genetic structure. We first consider PCA,
the most commonly used unsupervised method. Next, we investigate a sophisticated non--linear supervised classifier, a probabilistic ANN. Lastly we
consider a simpler but more limited linear supervised classifier, an SVM.

We would expect the supervised methods to perform better than PCA, since they utilise more information. The aim is to quantify this difference. We
therefore adopt a sliding window approach, using genetic windows of different sizes in order to to assess the perfomance of the classifiers given
different amounts of genetic data. 

According to a recent hypothesis, discussed below, unsupervised methods cannot distinguish between two populations if the amount of data available
falls below a certain threshold value. It is therefore of interest to determine whether supervised methods can classify below this  limit, and we
investigate this question also.

\subsection*{Principal Components Analysis}
\label{sec:pca}

The PCA technique is well known and commonly used in genetics and we do not describe it in detail here. Briefly, the aim is to determine the direction of
maximum variance in the space of data points. The first principal component points in the direction of maximum variance, the second component
maximises the remaining variance and so on. Any systematic difference between groups of individuals will manifest itself as a differential clustering
when the data points are projected on to these principal components.

We use the {\sc smartpca} component of the {\sc eigensoft} (v3.0) software package \cite{Patterson} for our analysis. In addition to the principal
components, {\sc smartpca} produces a biased but asymptotically consistent estimate of Wright's $F_{ST}$ parameter \cite{Reich2}. We use this
estimator as our measure of effect size.

The authors of {\sc smartpca} use a result obtained by \cite{Baik1} and \cite{Baik2}, to conjecture the existence of a phase transition (the Baik, Ben
Arous, P\'{e}ch\'{e} or BBP transition) below which population structure will be undetectable by PCA \cite{Patterson}. They further conjecture that
this threshold represents an absolute limit for \emph{any} (presumably unsupervised) classification method. For two populations of equal size, the
critical $F_{ST}$ threshold is given by:
$$ F_{ST}(crit) = { 1 \over \sqrt {N_{SNP}S}} $$
 where $N_{SNP}$ is the number of single nucleotide polymorphisms (SNPS) and $S$ is the total number of individuals in the dataset.
  
A measure of statistical significance between any pair of populations is also produced by {\sc smartpca}. This is obtained by computing the ANOVA
$F$-statistics for the difference in mean values along each principal component. A global statistic is calculated by summing over all components; this
statistic follows a $\chi^2$ distribution. We use the associated $p$-value as our measure of statistical significance.

It is important to point out that we are using the $p$-value as a quantitative measure.  This quantity is more usually used in a hypothesis testing
framework, where the decision to accept or reject is made on the basis of some pre-determined threshold. We do not set such a threshold; rather, we
use the $p$-value to detect the onset of the BBP phase transition, when its value drops by many orders of magnitude.

We determine the effectiveness or otherwise of PCA by comparing the estimated value of $F_{ST}$ with the critical value in a sliding window across the
chromosome.

\subsection*{Artificial Neural Networks}
\label{sec:mlp}

ANNs  are relatively uncommon in genetics and may be unfamiliar to many geneticists. Furthermore the network we employ possesses some novel
features particularly relevant to genetic analysis. We therefore give a somewhat more detailed overview in this section.

ANNs are a methodology for computing, based on massive parallelism and redundancy, features also found in animal brains. They consist of a number of
interconnected processors each of which processes  information and passes it to other processors in the network. Well-designed networks are
able to `learn' from a set of training data and to make predictions when presented with new, possibly incomplete, data. For an introduction to the
science of neural networks the reader is directed to \cite{Bailer}.

\begin{figure}
\begin{center}
\vspace{1cm}
\includegraphics[width=0.5\linewidth]{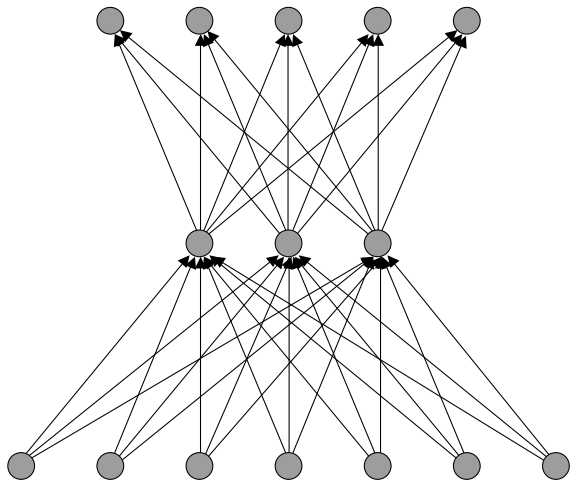}
\end{center}
\caption{\label{fig:nn} An example of a 3-layer neural network with
7 input nodes, 3 nodes in the hidden layer and 5 output nodes. Each line represents one weight.}
\end{figure}

The basic building block of an ANN is the \emph{neuron}. Information is passed as inputs to the neuron, which processes them and produces an
output.  The output is typically a simple mathematical function of the inputs. The power of the ANN comes from assembling many neurons into a
network. The network is able to model very complex behaviour from input to output. We use a three-layer network consisting of a layer of input
neurons, a layer of ``hidden'' neurons and a layer of output neurons.  In such an arrangement each neuron is referred to as a node.
Figure~\ref{fig:nn} shows a schematic design for this network with 7 input nodes, 3 hidden nodes and 5 output nodes.

The outputs of the hidden layer and the output layer are related to their inputs as follows:

\begin{eqnarray}
\mbox{hidden layer:} & h_j=g^{(1)}(f_j^{(1)}); &
f_j^{(1)} = \sum_l w^{(1)}_{jl}x_l +
  b_j^{(1)}, \\
\mbox{output layer:} & y_i=g^{(2)}(f_i^{(2)}); & f_i^{(2)} =
\sum_j w^{(2)}_{ij}h_j + b_i^{(2)},
\end{eqnarray}
where the output of the hidden layer $h$ and output layer $y$ are given for each hidden node $j$ and each output node $i$. The index $l$ runs over all
input nodes.  The functions $g^{(1)}$ and $g^{(2)}$ are called activation functions. The non-linear nature of $g^{(1)}$ is a key ingredient in
constructing a viable and practically useful network. This non-linear function must be bounded, smooth and monotonic; we use $g^{(1)} = \tanh x$. For
$g^{(2)}$ we simply use $g^{(2)}(x) = x$.  The layout and number of nodes are collectively termed the
\emph{architecture} of the network.

The weights $\mathbf{w}$ and biases $\mathbf{b}$ effectively define the network and are the quantities we wish to determine by some
\emph{training} algorithm. We denote $\mathbf{w}$ and $\mathbf{b}$ collectively by $\mathbf{a}$. As these parameters vary during training,
 a very wide range of non-linear mappings between inputs and outputs is possible. In fact, according to a `universal approximation
 theorem'~\cite{Leshno}, a standard three-layer feed-forward network can approximate any continuous function to \emph{any} degree of accuracy with
 appropriately chosen activation functions. However a network with a more complex architecture could well train more efficiently.

The use of ANNs in genetics to date has been limited.  A comprehensive review is given in \cite{motsinger2008}. Previous work has focused mainly on
investigating the optimum network architecture for specific applications, using a small number of genetic markers. A case-control scenario was
considered in \cite{curtis2007}. Their networks typically consisted of four input nodes, representing four markers, with two hidden layers
incorporating up to three hidden nodes each. The output was the case or control status of the individual. The authors explored a variety of different
architectures and assessed the performance of each.  In common with other authors such as \cite{north2003}, they noted that the performance of the
network was strongly dependent on the choice of architecture. Nevertheless, many authors such as \cite{seretti2004} and \cite{penco2005} have
successfully used ANNs with pragmatic choice of architecture based on trial and error searching.

A more serious  problem is the size of networks that it is possible to train when using traditional back-propagation or quasi-newtonian gradient descent
methods. Most such methods are very inefficient in navigating the weight space of a network and can therefore handle  only relatively small genetic
datasets.

Both these problems are addressed in the {\sc MemSys} package \cite{Gull} which we use to perform the network training. This package uses a
non--deterministic algorithm which allows us to make \emph{statistical} decisions on the appropriate classification. This makes possible the fast efficient
training of relatively large network structures on large data sets. Moreover the {\sc MemSys} package computes a statistic termed the Bayesian
evidence (see for example \cite{Jaynes} for a review). The evidence provides a mechanism for selecting the optimum number of nodes in the hidden layer
of our three--layer network.

We apply this ANN to our genetic classification problem by associating each input node with the value of a genetic marker from an individual and the
output nodes with the probabilities of the individual's membership of each class. As in the case of the PCA analysis we perform the classification in
a sliding window across the chromosome.

\subsection*{Support Vector Machines}
\label{sec:svm}

The ANN described in the previous section is a sophisticated classifier, able to amplify weak signals and to detect non--linear relationships in the
data. This feature is potentially of great significance in genetic analysis, since non--linearity is likely to arise due to long-range interactions
between genes at different physical locations. It is also of interest to investigate the performance of a more conventional linear supervised
classifier on the genetic classification problem. We therefore conduct a parallel analysis with an SVM.

\begin{figure}
\begin{center}
\psfrag{labelx}{$x_1$}
\psfrag{labely}{$x_2$}
\psfrag{label1}{$\zeta_1$}
\psfrag{label2}{$\zeta_2$}
\psfrag{label3}{$\zeta_3$}
\psfrag{label4}{$d_1$}
\psfrag{label5}{$d_2$}
\includegraphics[width=0.5\linewidth]{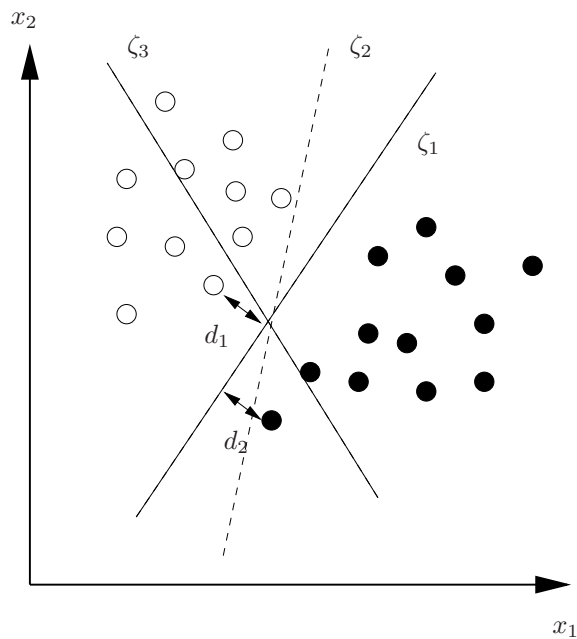}
\end{center}
\caption{\label{fig:svm} An example of a two-dimensional feature space $\mathbf{x}$
for data of known class divided by three hyperplanes $\zeta_i$. Clearly $\zeta_1$ divides most efficiently.}
\end{figure}

The principle of an SVM is intuitively very simple.  The space of data points  is partitioned by finding a hyperplane that places as many of the points
as possible into their pre-defined class.  The SVM algorithm iterates through trial planes, computing the shortest combined distance from the plane to
the closest of the data points in each class while simultaneously ensuring all data points of each class remain in the same partition. An example of a
two-dimensional feature space partitioned in three different ways is shown in Figure~\ref{fig:svm}.

In the example pictured the plane $\zeta_3$ does not partition the space correctly. The plane $\zeta_2$ produces an adequate classification with all
of the data points appropriately divided. However two data points lie very close to the plane and leave little margin for future
generalisation to unseen examples. The plane $\zeta_1$ is an optimum partitioning, maximising the combined distance $d_1 + d_2$. The function of an
SVM is to attempt to identify this optimum partition. In this work we  make use of the {\sc LIBSVM} library of SVM routines \cite{Chang}.

The SVM has the advantage of being simpler to use in practice, but has certain limitations compared with our ANN. Firstly it is a linear classifier
and cannot allow for non--linear relationships in the data. Secondly it is deterministic, providing  a unique solution for each problem. It is
therefore impossible to develop an estimate of the accuracy of the solution--that is, to place confidence limits on the classification. Our ANN, on
the other hand, is probabilistic, producing a slightly different solution on each iteration. This allows us to assess the stability of the
solution. Thirdly, the classification is binary--an individual either does, or does not, belong to a particular class. The ANN, in contrast, provides
probabilities of class membership for each class.

\subsection*{Data}
\label{sec:data}
Our test populations are  a subset of the data obtained by the International Schizophrenia Consortium (ISC). The consortium collected genome-wide
case--control data from seven sample collection sites across Europe. The final post quality controlled (QC) dataset contained 3322 cases and 3587
controls.  The controls from three sites were used for the purposes of this study:
\begin{itemize}
\item {\bf Aberdeen Site (P1)} A set of 702 controls, consisting of
volunteers recruited from general practices in Scotland. These were
genotyped on an Affymetrix 5.0 genotyping array.
\smallskip
\item {\bf Edinburgh Site (P2)} A set of 287 controls recruited through the
South of Scotland Blood Transfusion Service, typed on an Affymetrix
6.0 array.
\smallskip
\item {\bf Cardiff Site (P3)} A set of 611 controls recruited from several
sources in the two largest cities in Bulgaria, typed on an Affymetrix
6.0 array.
\end{itemize}
Quality control was performed by the ISC \cite{ISC2}. In addition to the usual genotype and sample QC procedures, attempts were made to resolve
technical differences arising from the different genotyping arrays used by the various ISC sites. A multi-dimensional scaling analysis was also
performed to detect population stratification and remove outliers from each population.

We start with the cleaned ISC data comprising 739,995 SNPs, all samples having a call rate $> 0.95$ and all SNPs having minor allele frequencies
$>0.01$, with population outlier identifiers removed \cite{ISC2}.  For the purposes of this study we examine a linkage-disequilibrium (LD) pruned set
of 5739 SNPs ($r^2 < 0.2$) on chromosome 1, selecting only those that were common to both the Affy 5.0 and Affy 6.0 platforms. {\sc plink} v1.06
\cite{Plink} software was used for this data reduction. The parameters of the three test populations are given in Supplementary Information, Table S1.

\section*{Results}
We first perform a principal components analysis (PCA) on the three populations to determine whether the populations can be distinguished using an
unsupervised learning approach. We then carry out both ANN and SVM supervised learning classifications on the same three populations.

\subsection*{PCA Classification}

We first test for structure \emph{within} each of our three populations. In each case the population is divided into two disjoint subsets. For P1 and
P3 each subset consists of 200 samples. In the case of P2, only 287 samples are available in total, so we divide these into two subsets of 140 samples
each. We do not remove any residual (post QC) outliers, in order to maximise any signal.

In all three cases we find that the estimated $F_{ST}$ values are vanishingly small, less than 0.0001 even when all 5739 SNPs are used. In no case do
the estimated levels of $F_{ST}$ exceed $F_{ST}(crit)$. By comparison a recent study \cite{Nelis} found values ranging as high as 0.023 across
Europe. The ANOVA $p$-values for the three populations P1, P2, and P3  are 0.050, 0.559 and 0.022 respectively. Although two of these $p$-values fall
at or below the conventional threshold of 0.05 this does not in itself imply the ability to detect structure in the absence of a reasonable effect
size. The PCA plot for the most significant case ($p$ = 0.022) shows that the populations do not separate ( Figure S1 , Supplementary Information). We conclude
that PCA fails to detect structure between the subsets tested in each of our three populations; that is, each population is essentially homogeneous.

\begin{figure}[p]
\begin{center}
\includegraphics[width=0.23\textwidth, angle=270]{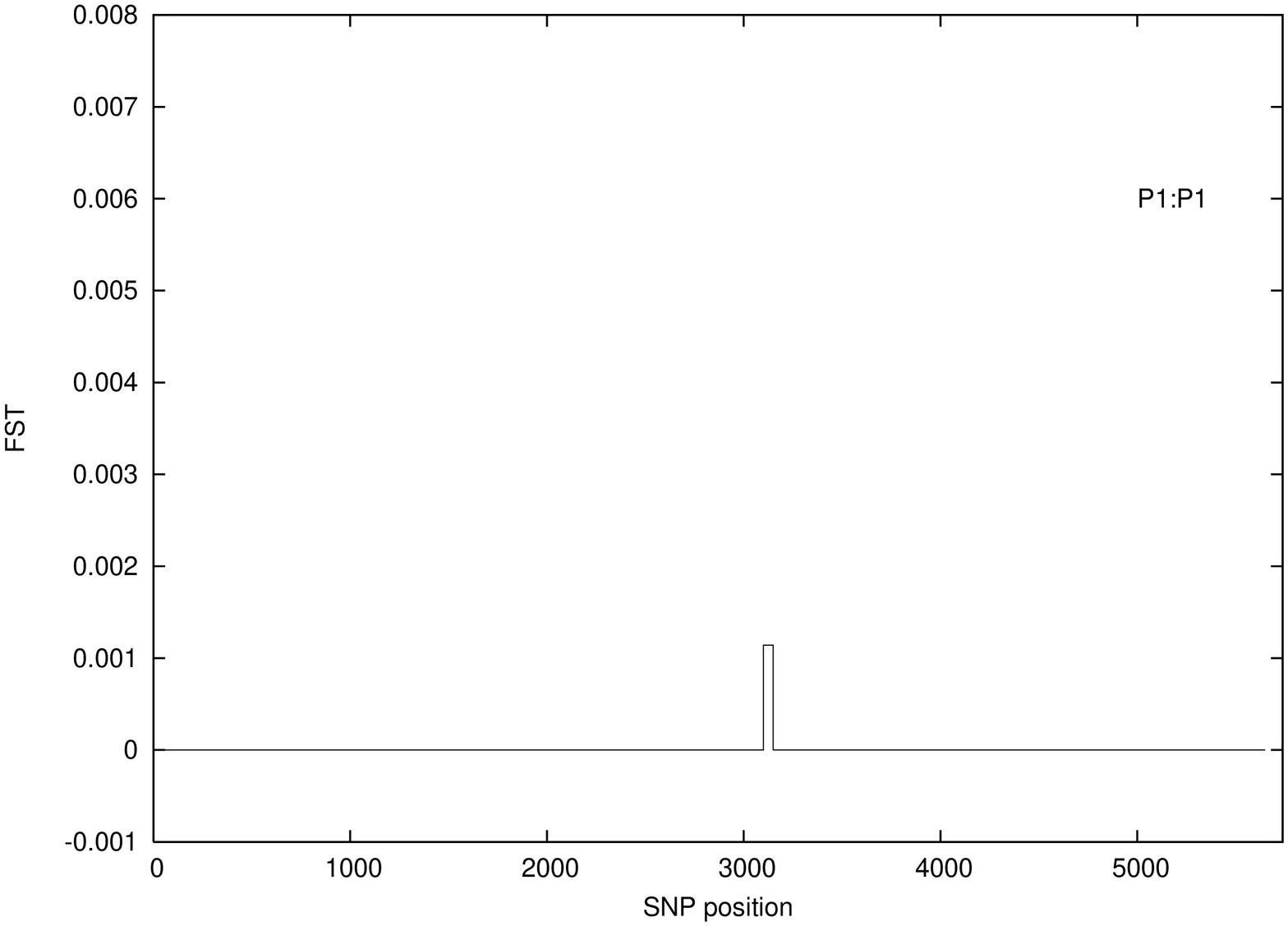}
\includegraphics[width=0.23\textwidth, angle=270]{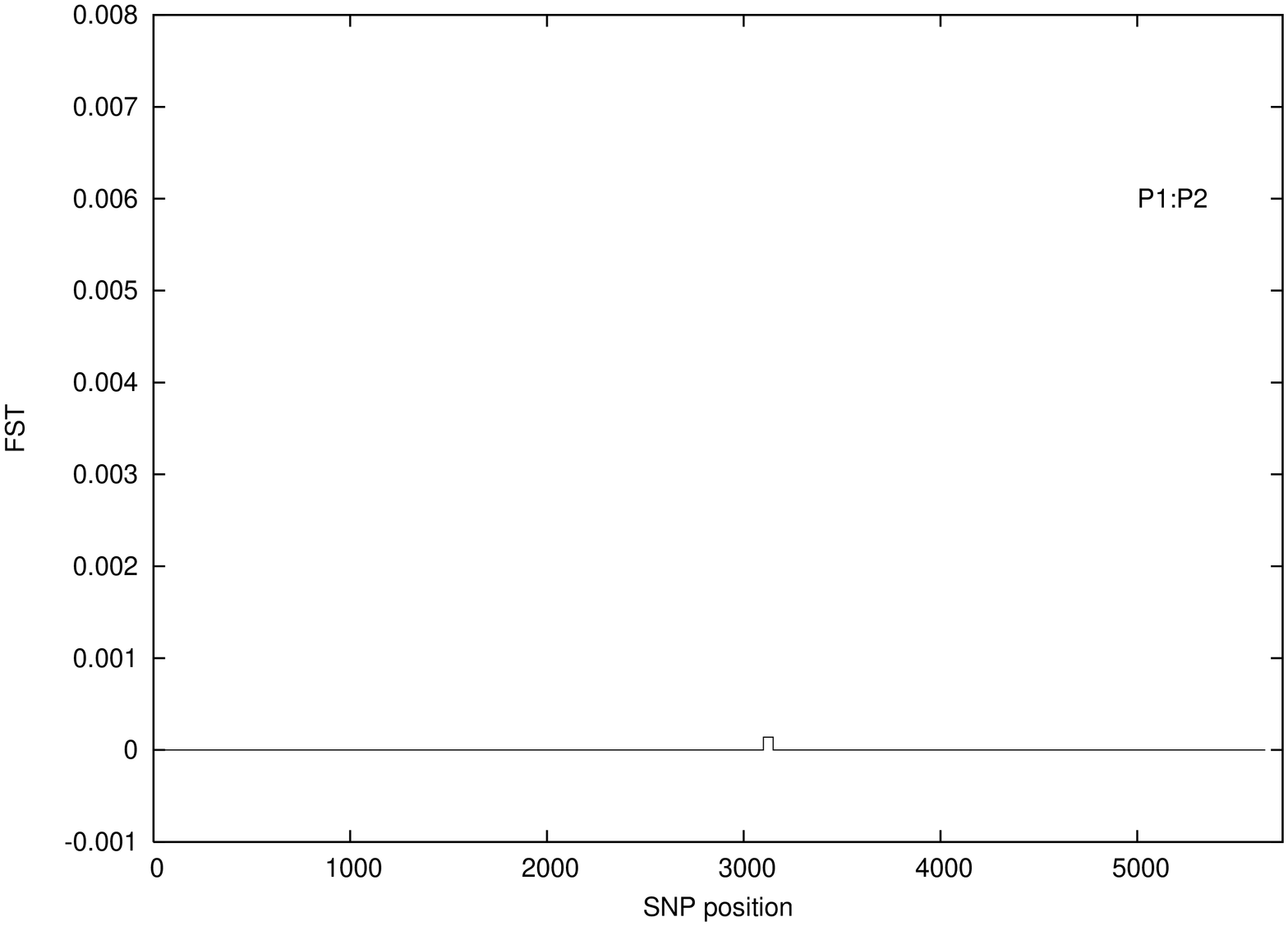}
\includegraphics[width=0.23\textwidth, angle=270]{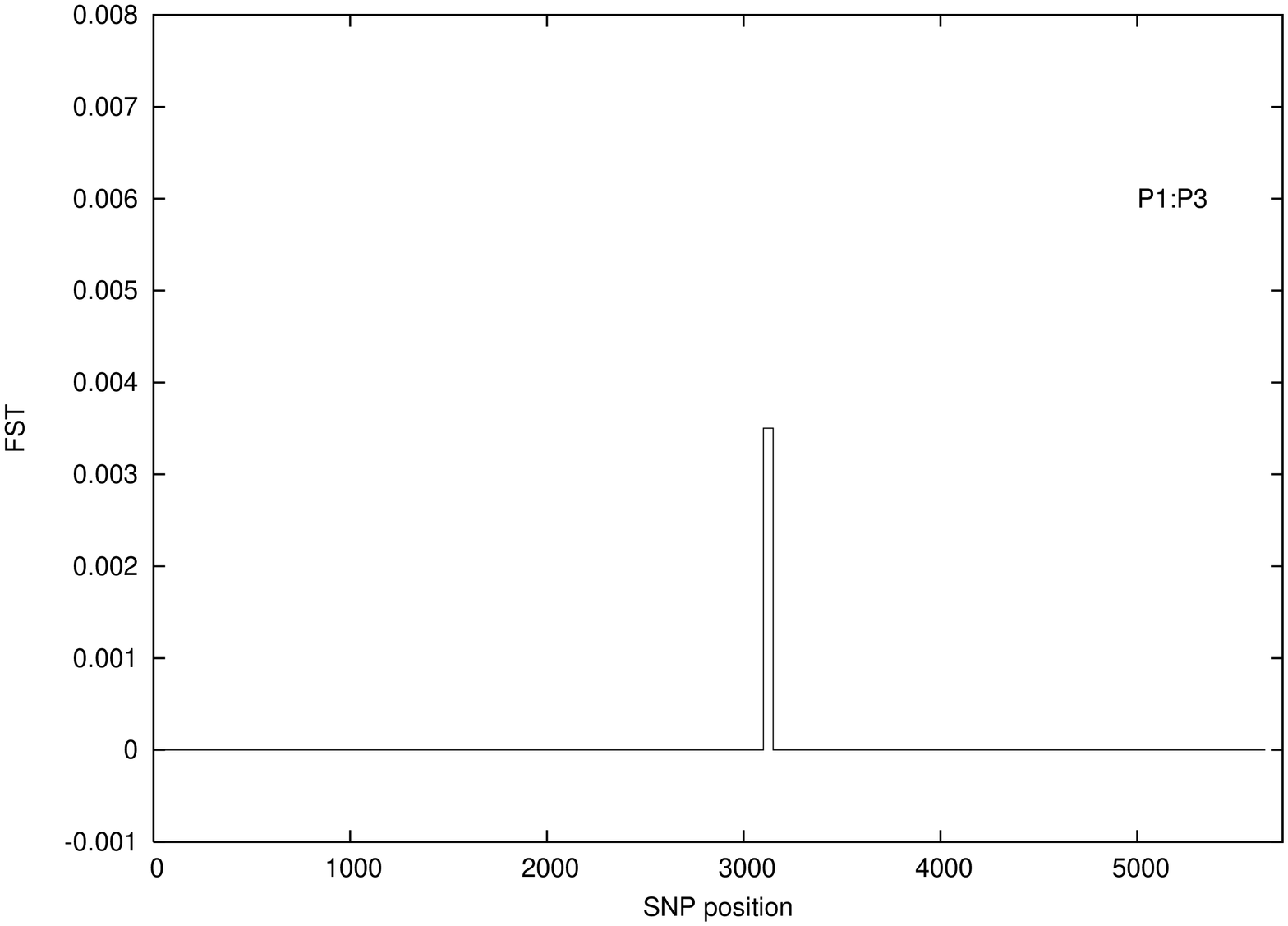}
\caption[50 SNP window]{Estimated $F_{ST}$ values  for 50 SNP sliding window. The
$F_{ST}$ is essentially zero everywhere except for a small region
approximately halfway along the chromosome. The horizontal dotted line
 is the value of $F_{ST}(crit)$}
\label{SW1}
\end{center}
\end{figure}

\begin{figure}[p]
\begin{center}
\includegraphics[width=0.23\textwidth, angle=270]{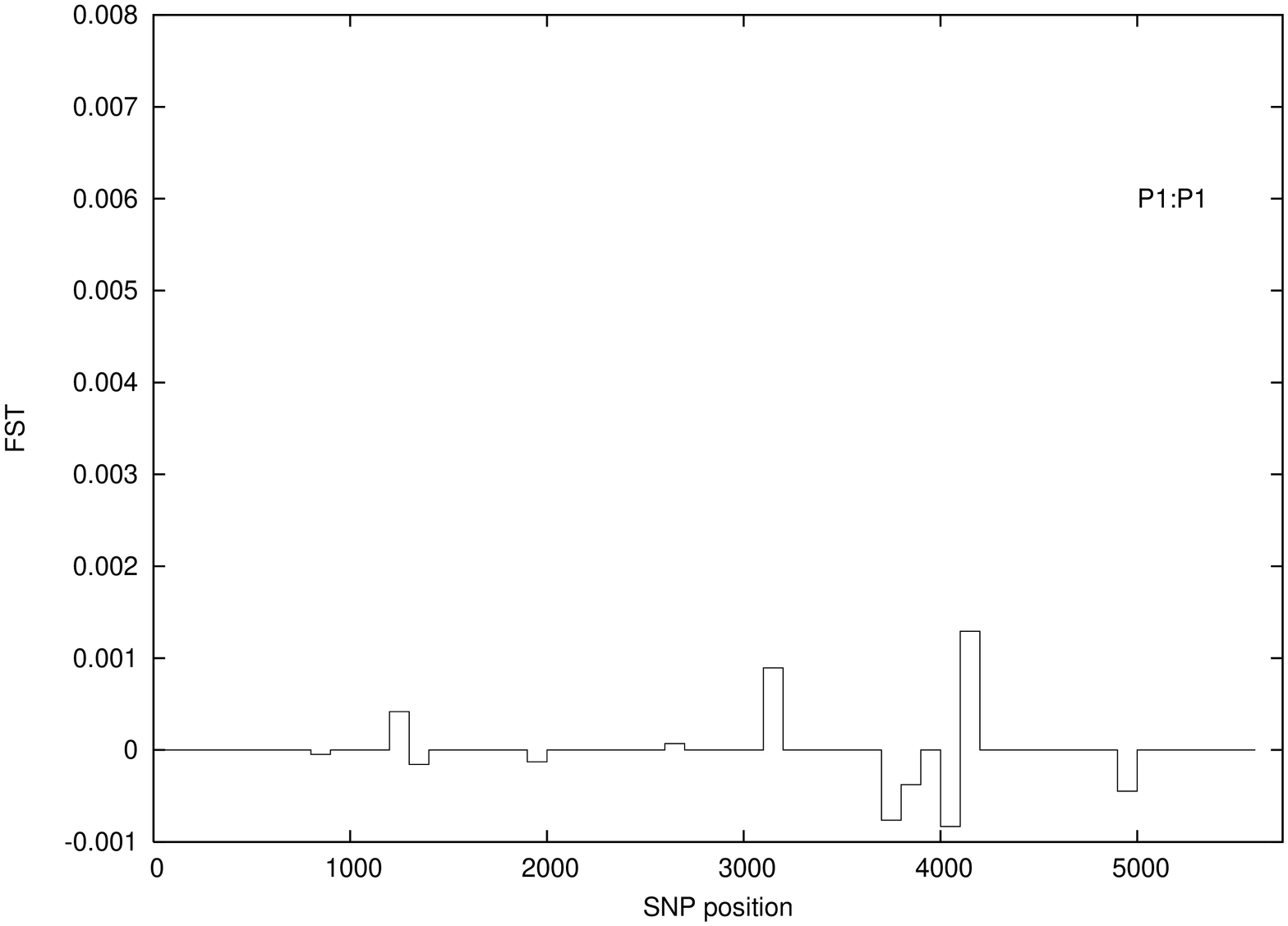}
\includegraphics[width=0.23\textwidth, angle=270]{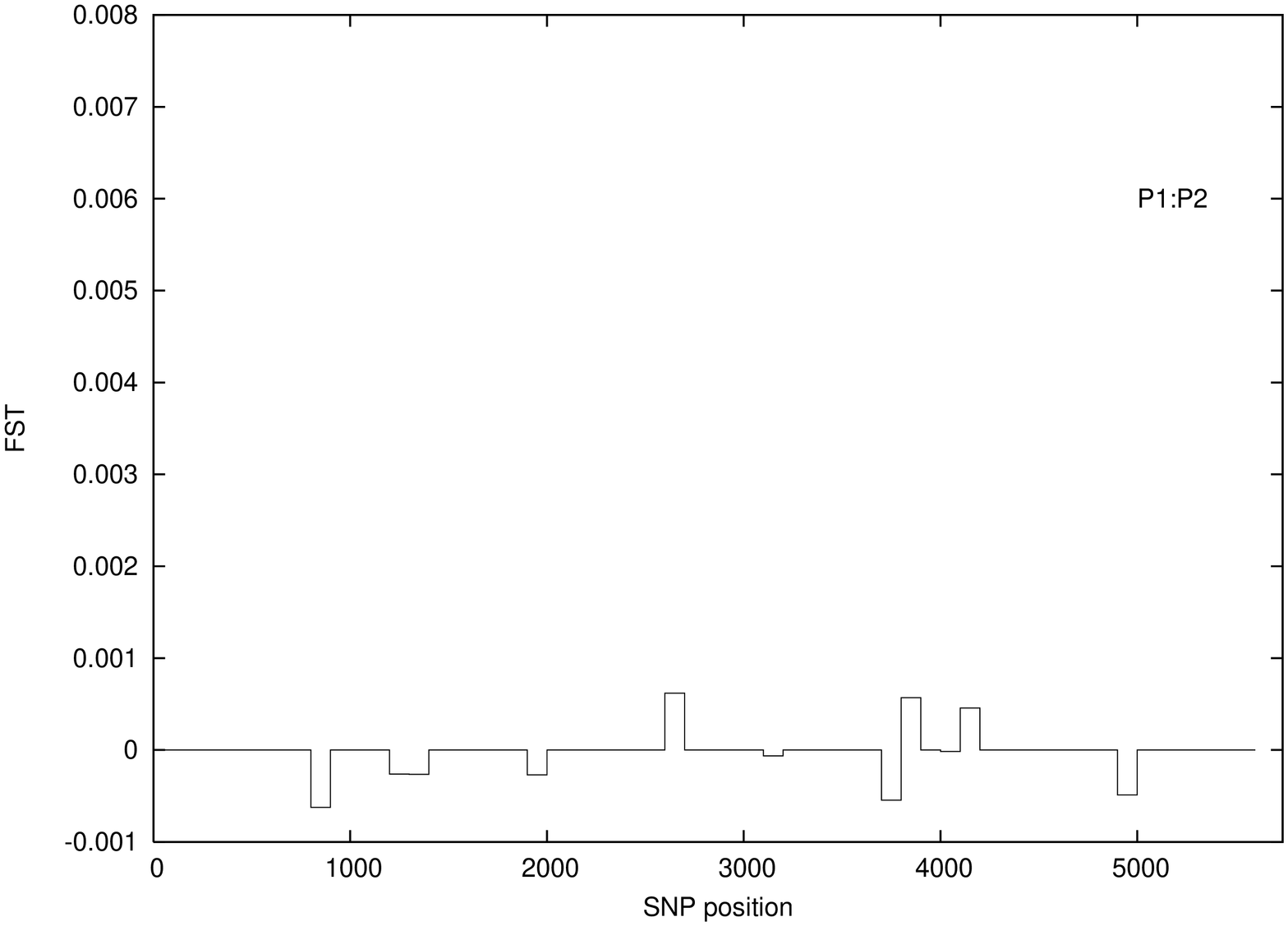}
\includegraphics[width=0.23\textwidth, angle=270]{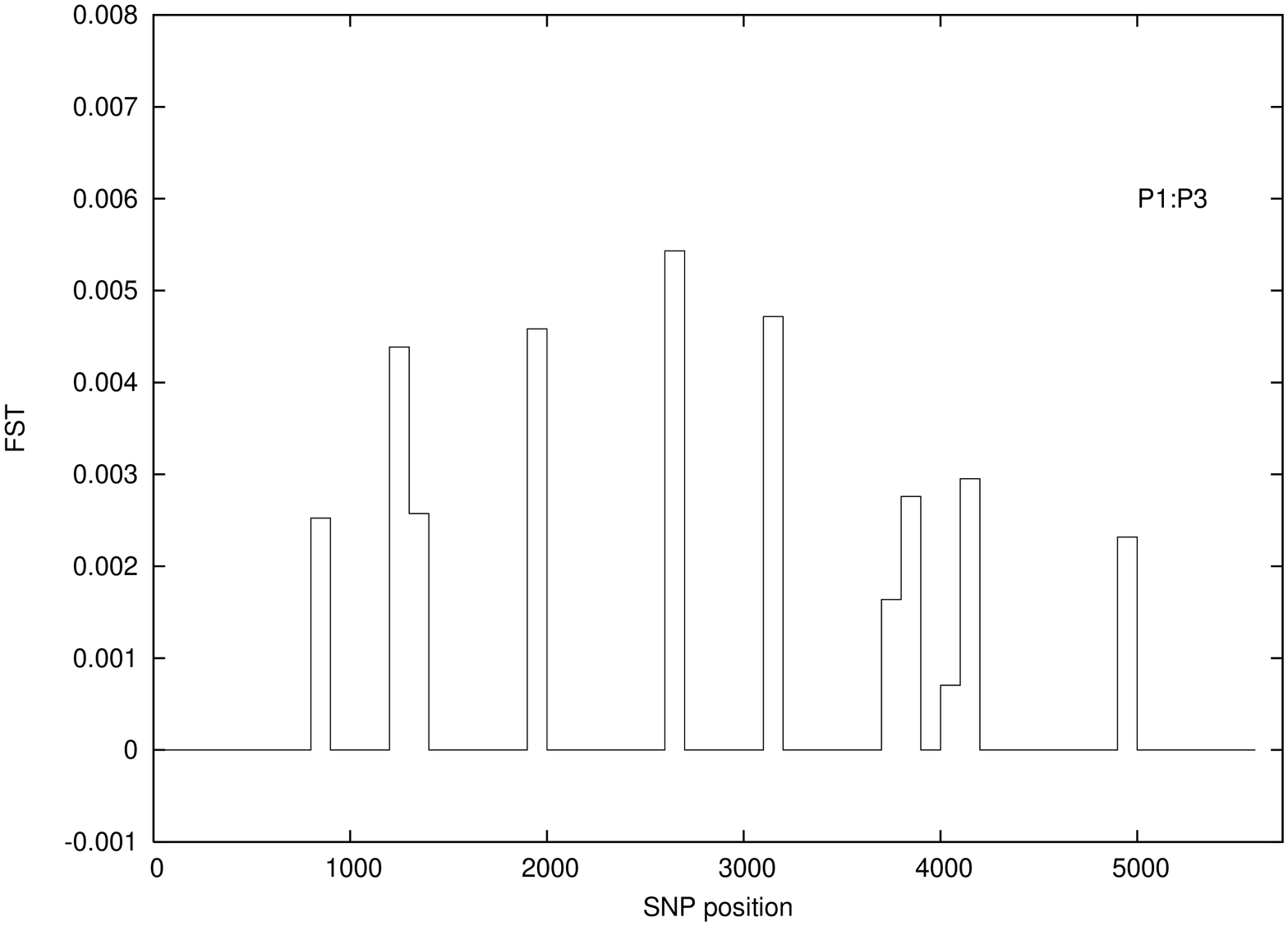}
\caption[100 SNP window]{Estimated $F_{ST}$ values for 100 SNP sliding window.
 The horizontal dotted line is the value of $F_{ST}(crit)$. The critical value is exceeded at only one location in the P1:P3 comparison. Note that
 although $F_{ST}$ is always non-negative, the estimator may become negative for small values of $F_{ST}$}
\label{SW2}
\end{center}
\end{figure}

\begin{figure}[p]
\begin{center}
\includegraphics[width=0.23\textwidth, angle=270]{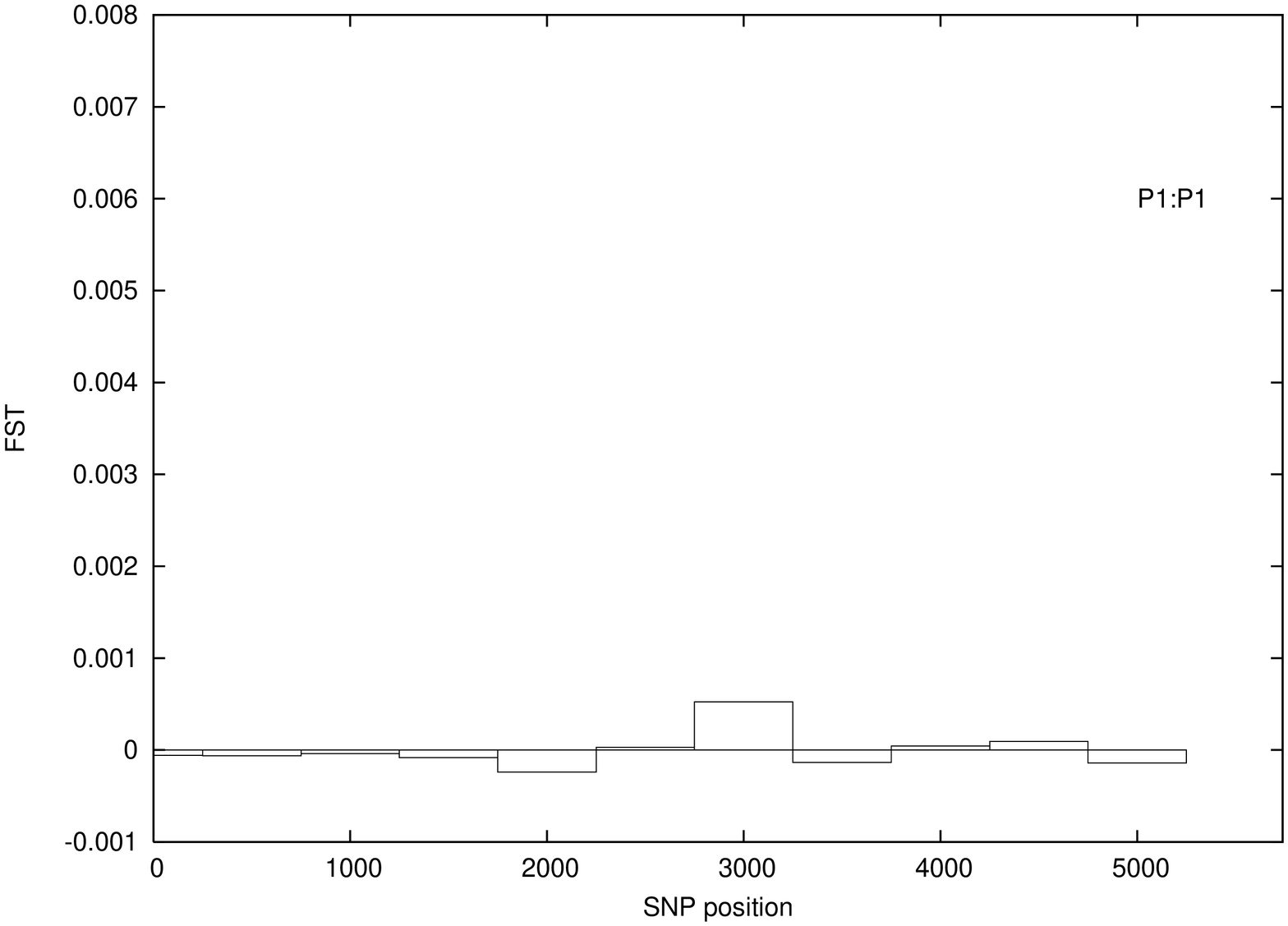}
\includegraphics[width=0.23\textwidth, angle=270]{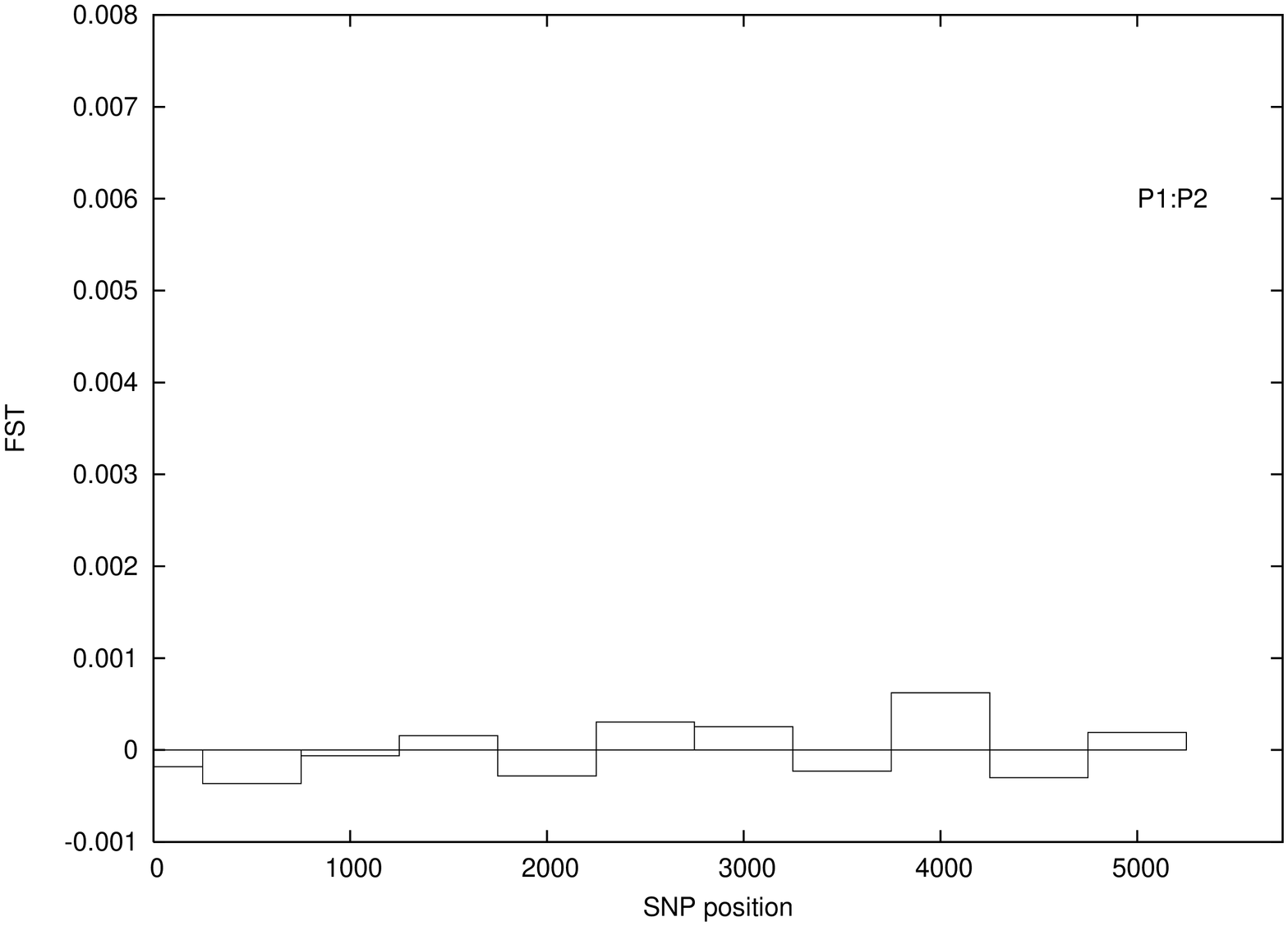}
\includegraphics[width=0.23\textwidth, angle=270]{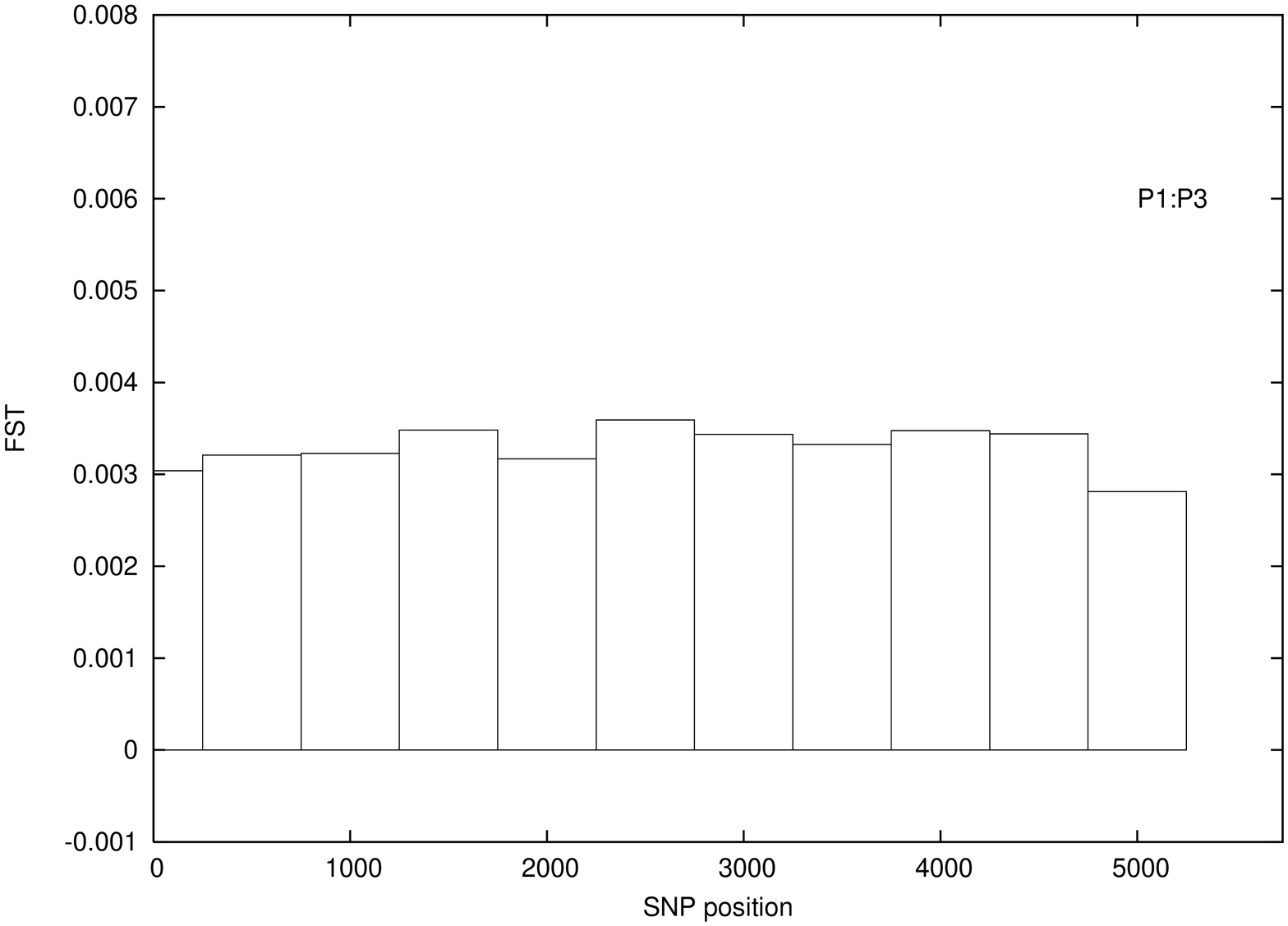}
\caption[500 SNP window]{Estimated $F_{ST}$ values for 500 SNP sliding window.
 The horizontal dotted line is the value of $F_{ST}(crit)$.  The
 analysis fails to distinguish between the P1 and P2 populations, but
 distinguishes between P1 and P3 everywhere along the chromosome.}
\label{SW3}
\end{center}
\end{figure}

 We next test for differences between our three populations. We perform a sliding window PCA analysis with non--overlapping windows of length 50, 100
 and 500 SNPs. The estimated $F_{ST}$ values are plotted in Figures \ref{SW1}, \ref{SW2} and \ref{SW3} with the corresponding critical value shown for
 comparison.


The estimated $F_{ST}$ is negligible at the 50 SNP level, except for one window about halfway along the chromosome, and even here it does not approach
$F_{ST}(crit)$. Some signals are visible for the P1:P3 comparison at the 100 SNP level, but $F_{ST}(crit)$ is exceeded in only one window.  At the 500
SNP level the PCA analysis can distinguish between the P1 and P3 populations, with the estimated $F_{ST}$ exceeding $F_{ST}(crit)$ everywhere along the chromosome but the P1:P2 comparison still shows negligible signal.  The full results from this analysis are given in Supplementary Information,
Table S2. Sample PCA plots showing the BBP transition given in Supplementary Information, Figures S2 and S3.

We may summarise the results of our PCA analysis as follows. As expected, no internal structure is detectable within any of the three
populations. Moreover, PCA is unable to distinguish the two Scottish populations even when using the full input set of 5739 SNPs. The two Scottish
populations can, however, be distinguished from the Bulgarian population, given an input data set of around 500 SNPs, anywhere along the chromosome.

\subsection*{ANN Classification} 
\label{sec:results_ann_svm}

We next attempt to classify the same data using the ANN. The pre-classified data available is divided into a \emph{training set} used to train the
network and a \emph{hold-out set} used to assess the accuracy of the network after training.  Since we merely wish to determine whether the ANN
is able to classify or not, it is desirable to to maximise the size of the training set while retaining a large enough testing set to ensure
statistically meaningful results. In practice we find that a ratio of  $80\%:20\%$ to be satisfactory and all the results presented here use this
ratio.

As with the PCA analysis we use samples of 200 from each population, except in the P2:P2 case, where we use 140 for each sub-population. We
perform multiple repetitions of the network training, drawing a different random starting point (of the weights and biases) on each occasion.  In this
way we are able to obtain an ensemble of trained classifiers from which we can draw a standard $1\sigma$ error on the network classification. For all
of the results below we use $>20$ repetitions. We present all of our results in terms of $\%$ accuracy of classification on the hold-out set, where
100$\%$ defines a perfect classifier and 50$\%$ is no better than random.

\begin{figure*}
\begin{center}
\psfrag{ylabel}{\tiny \% Classification}
\psfrag{xlabel}{\tiny SNP \#}
\includegraphics[width=0.35\linewidth, angle = -90]{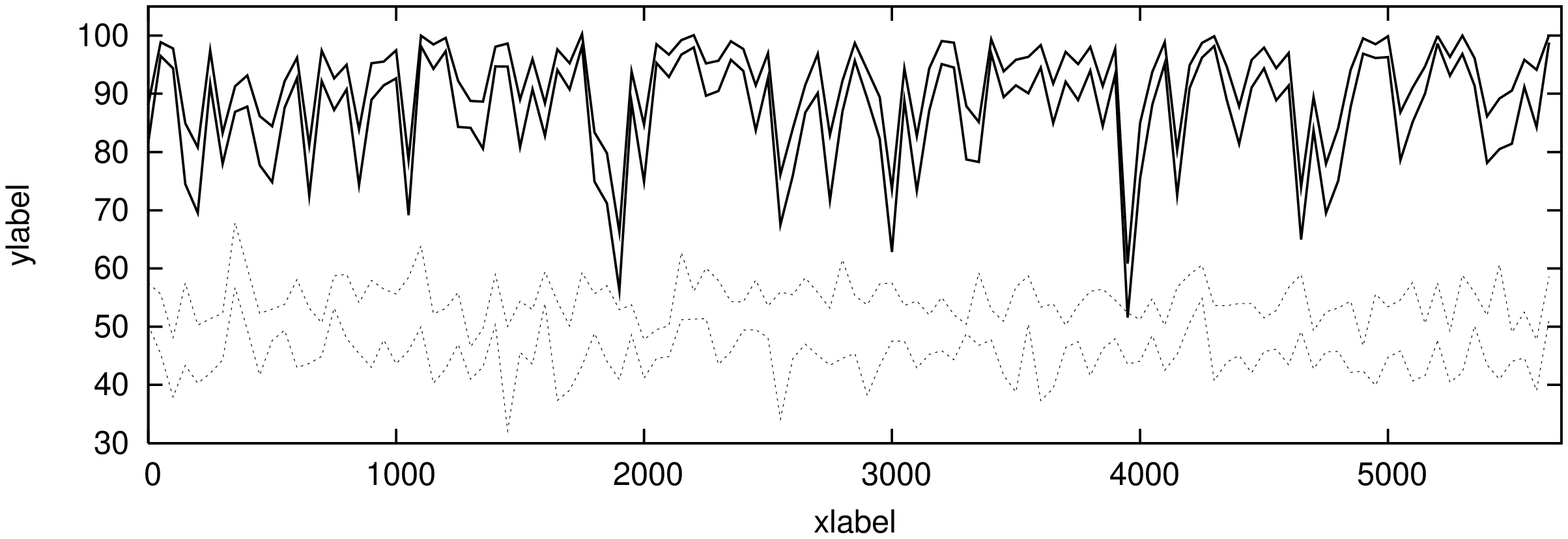}\\
\includegraphics[width=0.35\linewidth, angle = -90]{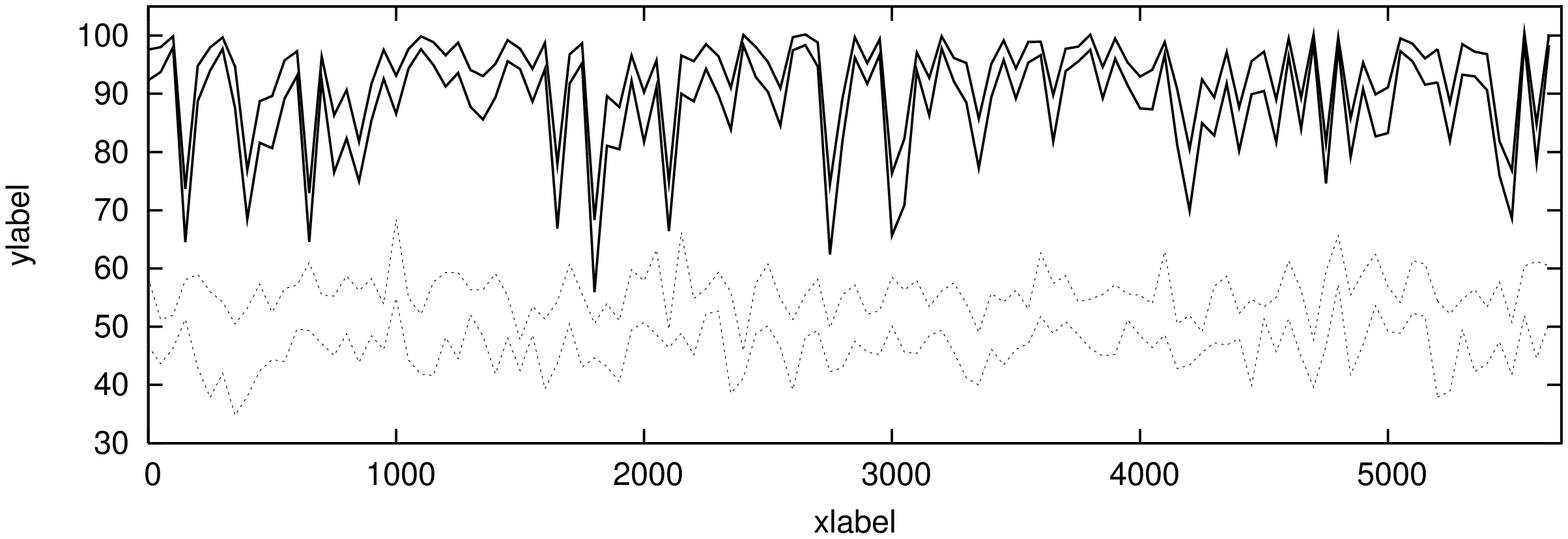}\\
\includegraphics[width=0.35\linewidth, angle = -90]{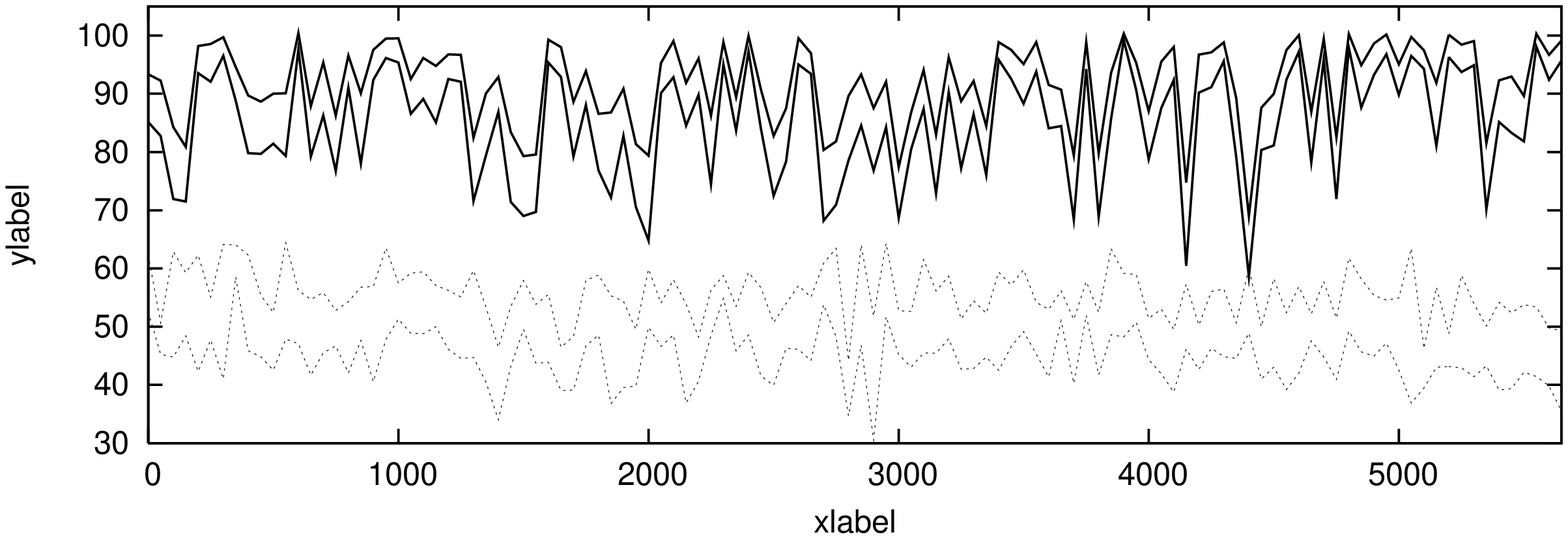}
\end{center}
 \caption{Top panel shows classification with windows of 50 contiguous, non-overlapping SNPs for P1 against P2 (solid lines) with classification
 results for a sample of P1 against P1 (dotted lines) shown for comparison. The regions enclosed between the lines illustrate 1$\sigma$ confidence
 intervals. The middle and bottom panels show the same results for P1 against P3 and P2 against P3 respectively.}
\label{fig:nn_fig1}
\end{figure*}

To explore the variation of classification across the chromosome we use an input set of non-overlapping windows each containing 50 SNPs. Figure
\ref{fig:nn_fig1} show the classification rate along the chromosome for each population combination. In addition each figure illustrates a reference
null classification of two sub-samples from each of the three populations to demonstrate the internal homogeneity of each population.

It is notable that a classification rate of $>80\%$ is achieved across the majority of the chromosome for \emph{both} populations P1:P2 \emph{and}
P1:P3. This demonstrates that the network can successfully amplify a much weaker, intra-Scottish population signal to roughly the same level as that
obtained for the Scotland-Bulgaria comparison.

\begin{figure*}
\begin{center}
\psfrag{ylabel}{\tiny \% Classification}
\psfrag{xlabel}{\tiny SNP \#}
\includegraphics[width=0.5\linewidth, angle = -90]{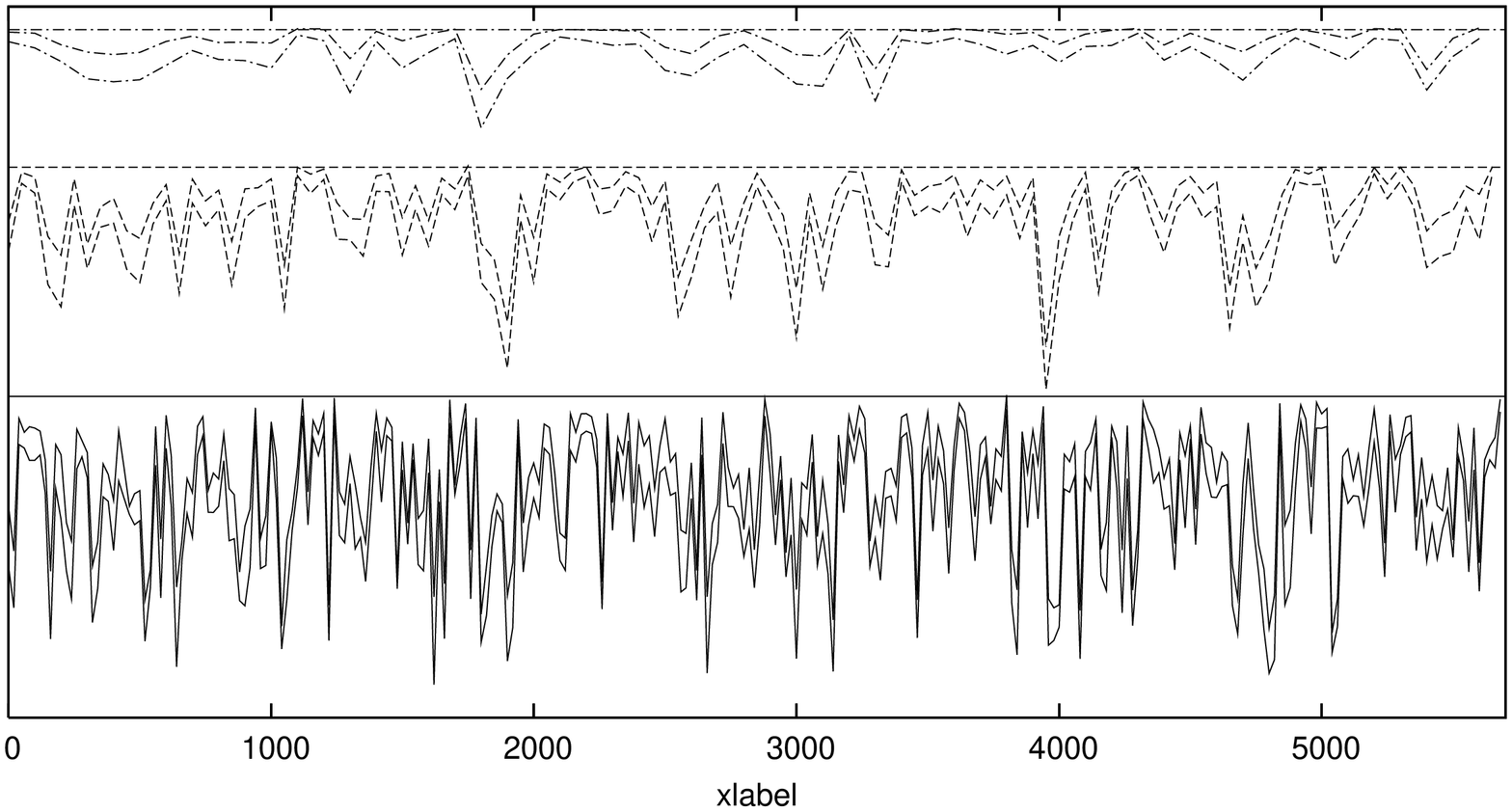}
\end{center}
 \caption{Classification with windows of 100 (dot-dashed), 50 (dashed) and 20 (solid) contiguous, non-overlapping SNPs for P1 against P2.  Note that
 as the window size increases, the accuracy converges to the \emph{most} accurate classification, indicating that the ANN is successfully discarding
 irrelevant information.  For clarity we have added an offset to each spectrum and omitted the ordinate axis, the horizontal lines represent $100 \%$
 classification in each case. }
\label{fig:nn_fig2}
\end{figure*}

We next investigate the variation in performance as the window size is varied.  Figure \ref{fig:nn_fig2} shows results for the classification of P1:P2
 with window sizes of 20, 50 and 100 SNPs. 
For a window size of 20, one sees considerable structure along the chromosome, with some regions classifying well, and others poorly.  As the window
size increases, with each window now containing both ``good'' and ``bad'' regions, we find that the classification rate converges to the best,
rather than the worst rate. This shows that even when the network is presented with a large window that contains a small proportion of
informative SNPs it can successfully filter out the extraneous inputs and produce a classifier with the same level of accuracy as would have been
obtained with a reduced set of informative inputs. This feature has many important implications within genetics where data is often noisy or
incomplete.

\begin{figure} 
\begin{center}
\psfrag{xlabel}{\tiny FPR}
\psfrag{ylabel}{\tiny TPR}
\includegraphics[width=0.33\linewidth, angle =-90]{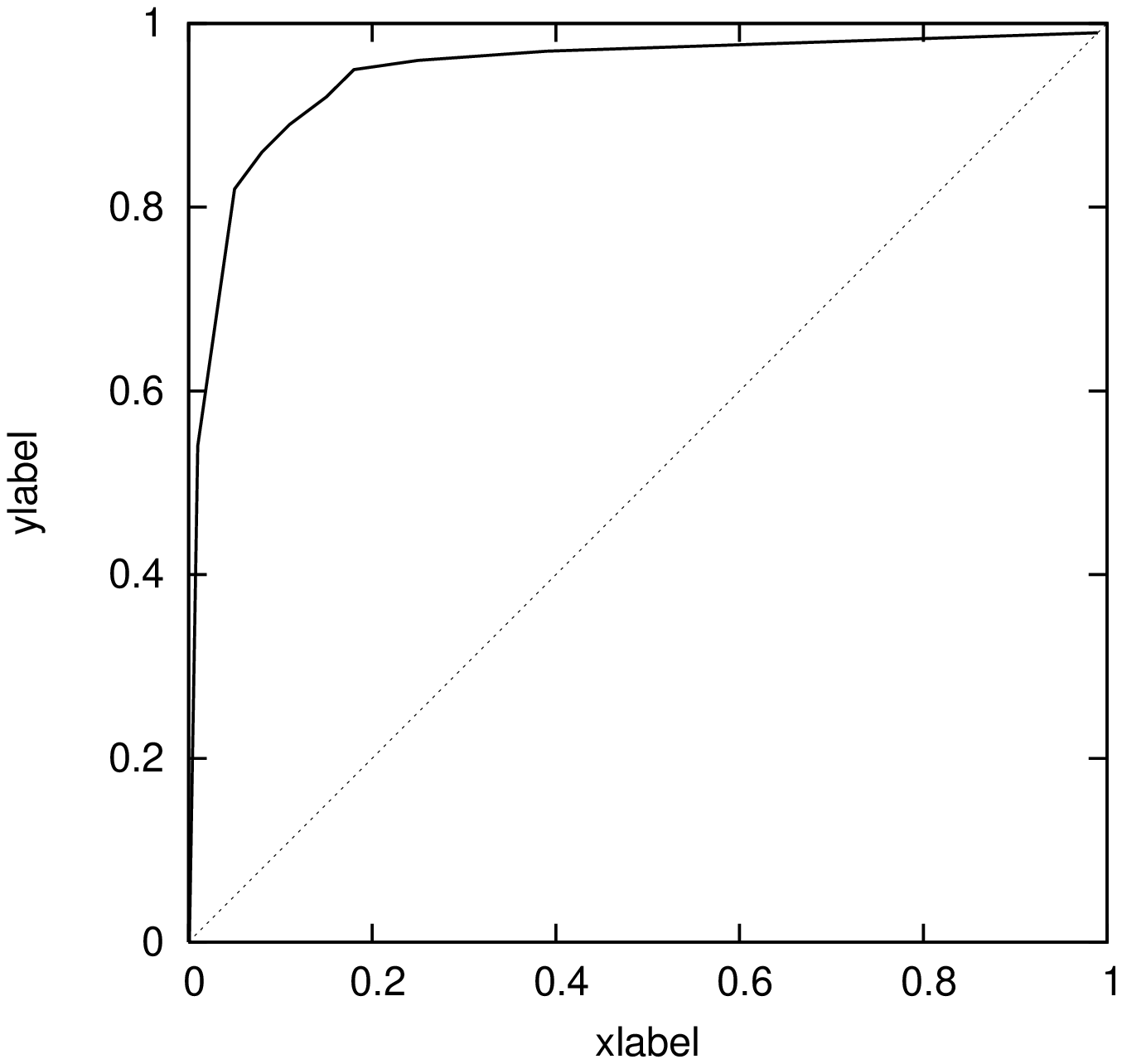}
\includegraphics[width=0.33\linewidth, angle =-90]{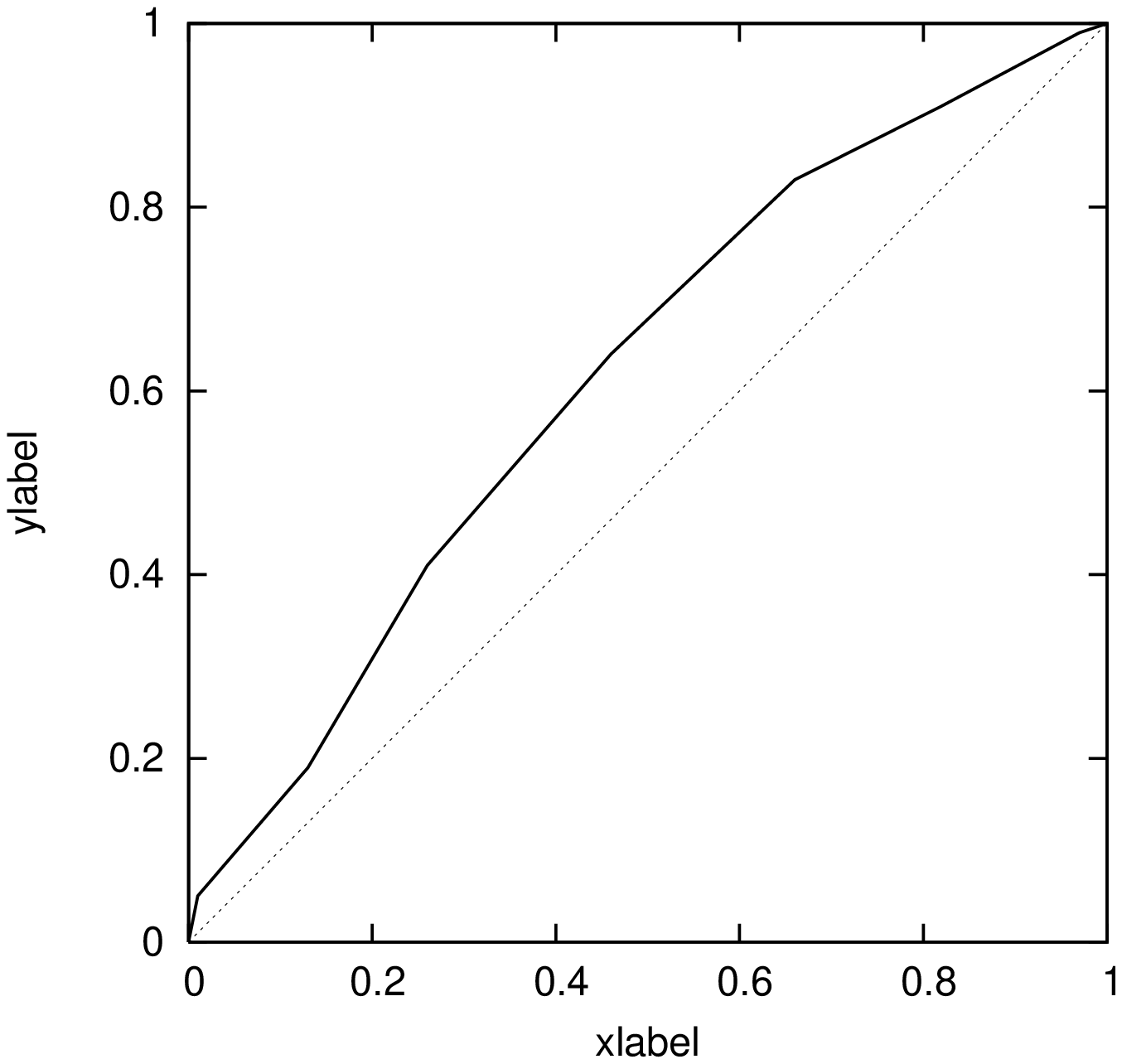}
\end{center}
\caption{\label{fig:nn_classification_ROC} 
Left panel shows a Receiver Operating Characteristic (ROC)
curve, that is a plot of true positive rate (TPR) against false positive rate (FPR) 
of the neural network classifier trained using the first 50 SNPs using P1 and P2
(solid curve). A random classifier (dotted curve) is shown for comparison.
Right panel shows the same for a classifier trained using a window of SNPs from 1950 to
2000.   
}
\end{figure}

It is common in signal processing to represent the efficiency of a classifier graphically, using a receiver operating characteristic (ROC) curve which
plots the true positive rate (TPR) versus the false positive rate (FPR) for increments of the classifier's discrimination threshold. The default
threshold is normally $0.5$, but variation of this criterion allows classifiers to be tuned to minimise the FPR while simultaneously maximising the
TPR. An ideal classifier has a ROC curve that resembles a step-function with a TPR of $1.0$ for all values of the threshold, while the ROC curve for a
random classifier is a line with slope of unity from a TPR of $0$ to $1$. Figure \ref{fig:nn_classification_ROC} illustrates the ROC curves for the
network classifier in two different regimes along the chromosome spectrum.  The left panel shows the ROC curve of the classifier trained using the
first 50 SNPs. As is evident from Figure~\ref{fig:nn_fig1} this region produces a classifier that is capable of distinguishing the two population
groups at the $90\%$ level. The quality of this classifier is then clearly discernible by a ROC curve that approaches a step-function. For comparison
we performed the same test on a part of the chromosome spectrum where the classifier was relatively poor, at a SNP window of $1950-2000$. This ROC
curve, shown in the right panel of Figure~\ref{fig:nn_classification_ROC} appears very close to the random classifier line, as would be
expected. Along with multiple network realisations computed for each classifier these tests provide a useful way to confirm the stability of the
classifiers.

The architecture of our three layer network is determined entirely by the number of nodes in the hidden layer. This number in turn can be estimated
from the Bayesian evidence. We find that our results are insensitive to the number of hidden nodes. In fact, reducing the number of hidden nodes from 20
to zero results in negligible degradation in performance, indicating that the signal we detect is essentially linear.  It is of course possible to
identify such a linear signal using PCA for example, given a signal of sufficient strength, as was demonstrated in the earlier part of this paper. The
reason for the increased sensitivity of our ANN here is its utilisation of our prior knowledge of class membership and its efficiency in exploring the
space of all \emph{possible} linear (and non-linear) mappings and identifying the choice that maximises the classifier's sensitivity automatically.

In summary, we find that the ANN exhibits considerably greater sensitivity than PCA. In particular, while PCA cannot distinguish between the two
Scottish populations, the ANN can do so given fewer than 100 SNPs. Moreover, the ANN can classify on a dataset well below the BBP limit. Furthermore,
as we have seen, the ANN can also efficiently eliminate noise. Our results indicate that the signal the ANN is identifying is linear, but nevertheless
too weak for PCA to detect.

\subsection*{SVM Classification}
\label{sec:results_svm}

\begin{figure*}
\begin{center}
\psfrag{ylabel}{\tiny \% Classification}
\psfrag{xlabel}{\tiny SNP \#}
\includegraphics[width=0.35\linewidth, angle = -90]{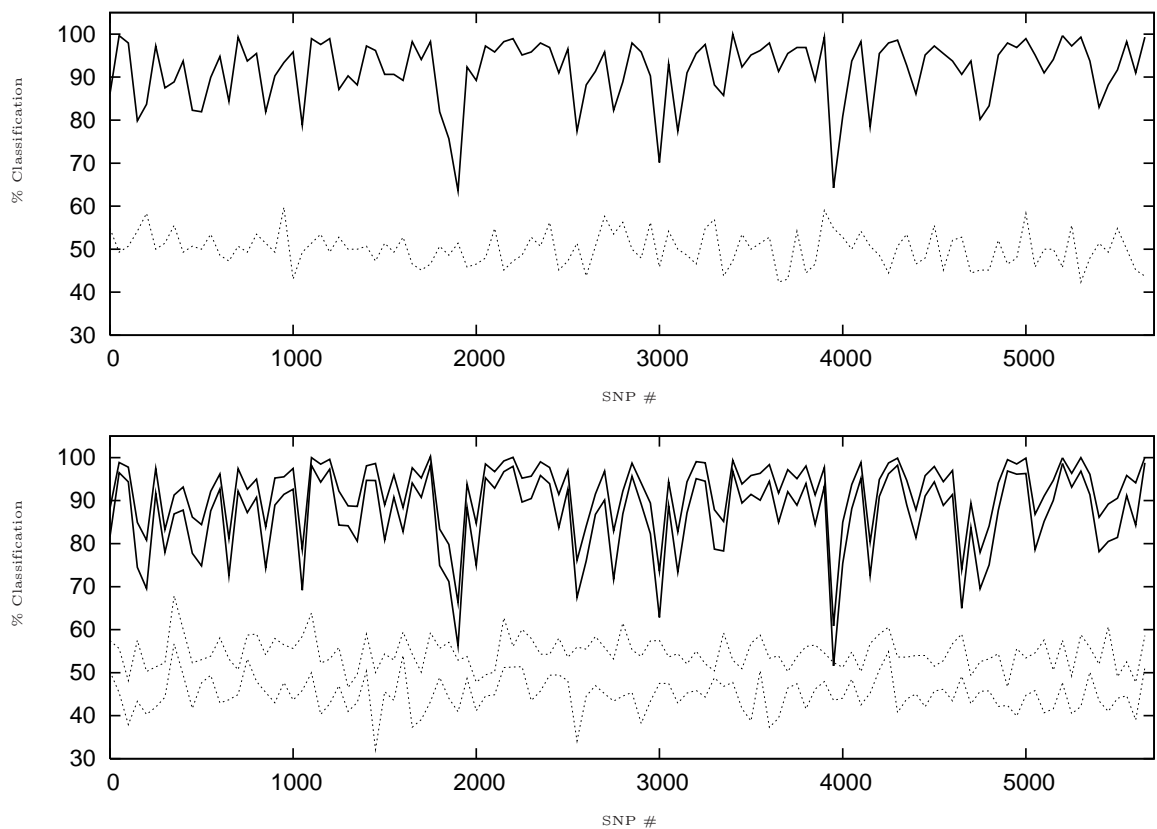}\\
\includegraphics[width=0.35\linewidth, angle = -90]{PLOTS/P1_P2_ANN.eps}
\end{center}
\caption{Top panel shows SVM classification with windows of 50 contiguous, non-overlapping SNPs
 for P1 against P2 
(solid lines) with classification results for a sample of  P1 against P1 (dotted lines) 
shown for comparison. Bottom panel shows the equivalent analysis with the ANN.}
\label{fig:results_svm}
\end{figure*}

In view of the fact that the dominating signal in the data is linear, we would expect the SVM to perform equivalently. We do not repeat the entire
analysis here, but simply show the sliding window analysis for the population combination P1 and P2 in Figure~\ref{fig:results_svm} (with the
equivalent ANN results for comparison). Since the SVM for a given dataset is entirely deterministic it is not possible to generate multiple
realisations of the classifier and thus build up $1\sigma$ confidence intervals. However it is clear that SVM performs comparably with the ANN on this
dataset,  locating strikingly similar features in the classification spectrum across the chromosome. It is also of interest to compare  the speed of each
method. The SVM takes roughly 10 seconds to build a classifier on a  $50$ SNP window,  using a currently standard desktop computer. A single iteration
of the ANN takes a roughly equal amount of time, with $1\sigma$ limits being generated in a $n_{\rm iterations}$ multiple of this time.
\label{sec:results_svm}

\section*{Discussion} 
\label{sec:discuss}

We demonstrate in this paper that supervised learning classification is to be preferred to unsupervised learning in genetics, when we have an \emph {a
priori} definition of class membership from some non-genetic source. The classification then serves to determine whether or not the pre-defined
populations are \emph{genetically} distinguishable.

Both the techniques investigated in this paper (SVMs and ANNs) significantly outperform PCA on the data presented here. It is noteworthy that the
sensitivity of these methods exceeds the conjectured BBP limit on the sensitivity of supervised approaches.

Although ANNs have been previously discussed in the context of genetics, they have yet to come into common use in this field. This is probably due, in
part, to the limited number of input nodes that it was possible to handle, and in part to the difficulty of determining the optimal network
architecture. Our ANN allows us to handle very large numbers of inputs, an essential feature in many applications in genetics.  The problem of
deciding on the optimal network architecture, much discussed by previous authors, reduces, in the case of a 3-layer network, to deciding on the number
of hidden nodes; the {\sc MemSys} package provides a rigorous method of determining this number.

In the event, we observe a predominantly \emph{linear} signal on this dataset, easily detectable by both SVM and ANN but too weak to be detected by
PCA. In a sense, this is be expected, since the SVM and ANN utilise our prior knowledge of class membership to find the optimal linear mapping for
classifying the data. In the absence of such prior information, PCA finds the linear mapping that maximises the variance; this is not necessarily the
optimal mapping. However the sensitivity of the supervised methods and the small number of SNPs that they need in order to classify efficiently is
noteworthy. A further important consequence of this fact is that the SVM and ANN can \emph{localise} the sources of genetic difference along the
chromosome and indeed the results of both methods are consistent with each other in this respect.

The linearity of the signal means that the SVM and ANN perform comparably. (The main novelty here is the large number of inputs that our ANN can
accept). This linearity is not altogether surprising, since non--linear effects would arise as a result of long--range correlation between loci. The
relatively small size of our SNP windows greatly reduces the probability of seeing such correlations. (Short range correlations, which arise from
linkage disequilibrium, carry no useful information and were eliminated by LD pruning our data).

When a linear signal is present, both the ANN and the SVM can classify with equal efficiency and we recommend that both be considered for use in
genetic classification. The ANN, however, possesses three advantages over the SVM. Firstly the stochastic nature of the classification means that we
can place confidence limits on our results. Secondly, the ANN supplies explicit probabilities for the classification of each individual. This provides
the potential to ``clean'' our datasets by removing those individuals who classify with very high (or very low) probability. Thirdly, the ANN is capable
of being applied to more general datasets where non--linear signals are significant.

It is noteworthy that the supervised learning methods are able to classify individuals from two populations within Scotland. One would expect
sufficient gene flow to occur within this region to homogenise the populations. The differences detected are not necessarily due to ancestry, but may
be a consequence of the fact that the two population samples were drawn from different datasets, genotyped on different platforms, at different
sites. These differences, whatever their origin, are nevertheless too small to detect using PCA, but in many applications the presence of such
differences may be of critical importance.

The behaviour of our ANN in the presence of significant non--linear effects remains to be investigated; one possible target is the common disease
common variant (CDCV) model of complex diseases. These are associated with many common genetic variants, each of individually small
effect. Interactions between these variants are likely to result in non--linear effects suitable for study with ANNs.

We suggest, on the basis of the evidence presented in this paper, that supervised learning methods have a useful role to play in genetic applications
where we are interested in differences between pre--defined groups of individuals. Possible applications include population genetics, case--control
studies and quality control for genetic data gathered at different sites or on different platforms.

\section*{ Supporting Information}

{\bf Software}
\medskip
 
The {\sc LIBSVM} library of SVM routines is publicly available \cite{Chang}.  The {\sc MemSys} algorithms can be made available for academic use.  We
have developed an interface to both the {\sc MemSys} and  {\sc LIBSVM} packages for our specific genetic application and are currently developing it for more
general applications. We would be happy to collaborate with interested parties to facilitate this development process.

\bigskip

{\bf Members of the International Schizophrenia Consortium}
\medskip

{\bf Trinity College Dublin} Derek W. Morris, Colm
O'Dushlaine, Elaine Kenny, Emma M. Quinn, Michael Gill, Aiden
Corvin;

{\bf Cardiff University} Michael C.O'Donovan, George K. Kirov,
Nick J. Craddock, Peter A. Holmans, Nigel M.Williams, Lucy Georgieva,
Ivan Nikolov, N. Norton, H. Williams, Draga Toncheva,Vihra Milanova,
Michael J. Owen; 

{\bf Karolinska Institutet/University of North
Carolina at Chapel Hill} Christina M. Hultman, Paul Lichtenstein, Emma
F.Thelander, Patrick Sullivan; 

{\bf University College London} Andrew
McQuillin, Khalid Choudhury, Susmita Datta, Jonathan Pimm, Srinivasa
Thirumalai, Vinay Puri, Robert Krasucki, Jacob Lawrence, Digby
Quested, Nicholas Bass, Hugh Gurling; 

{\bf University of Aberdeen}
Caroline Crombie, Gillian Fraser, Soh Leh Kuan, Nicholas Walker, David
St Clair; 

{\bf University of Edinburgh} Douglas H. R. Blackwood,
Walter J. Muir, Kevin A. McGhee, Ben Pickard, Pat Malloy, Alan
W. Maclean, Margaret Van Beck; 

{\bf Queensland Institute of Medical
Research} Naomi R. Wray, Peter M. Visscher, Stuart Macgregor;

{\bf University of Southern California} Michele T. Pato, Helena Medeiros,
Frank Middleton, Celia Carvalho, Christopher Morley, AymanFanous,
David Conti, James A. Knowles, Carlos Paz Ferreira, AntonioMacedo,
M. Helena Azevedo, Carlos N. Pato; 

{\bf Massachusetts General
Hospital} Jennifer L. Stone, Douglas M. Ruderfer, Manuel
A. R. Ferreira, 

{\bf Stanley Center for Psychiatric Research and Broad
Institute of MIT and Harvard} Shaun M. Purcell, Jennifer L. Stone,
Kimberly Chambert, Douglas M. Ruderfer, Finny Kuruvilla, Stacey
B. Gabriel, Kristin Ardlie, Mark J. Daly, Edward M. Scolnick, Pamela
Sklar.
\\
\bigskip

\section*{Acknowledgements}
We thank the individuals and families who contributed data to the International Schizophrenia Consortium. We are grateful to the reviewers for their
constructive comments. We thank the members of the Statistical Genetics Unit in the Neuropsychiatric Genetics Group for helpful comments and advice at
all stages of this work. We acknowledge Anthony Ryan for reviewing and commenting on this manuscript. The authors would like to acknowledge support
from the Cambridge Centre for High Performance Computing where this work was carried out, and also to Stuart Rankin for computational
assistance. Additionally we acknowledge Steve Gull for useful discussions and for the use of {\sc MemSys} in this application.

\medskip

\bibliographystyle{plos2009}
\bibliography{genpaper}

\end{document}


\center{\section*{Supplementary Information}}

\bigskip\bigskip
\bigskip\bigskip

\begin{table*}[!ht]
\begin{center}
\begin{tabular}{|c|c|c|c|} \hline
&&&\\
 & {\bf P1} & { \bf P2} & {\bf P3} \\
&&&\\
\hline\hline
&&&\\
{\bf ISC Site} & Aberdeen &  Edinburgh & Cardiff \\
&&&\\
\hline
&&&\\
{\bf Ancestry} & Scotland &  South East Scotland & Bulgaria \\
&&&\\
\hline
&&&\\
{\bf Sample Size} & 702 &  287 & 611 \\
&&&\\
\hline
&&&\\
{\bf SNPs} & 5739 &  5739 & 5739 \\
&&&\\
\hline
\hline
&&&\\
{\bf Array} & Affymetrix 5.0 &  Affymetrix 6.0 & Affymetrix  6.0\\
&&&\\
\hline
&&&\\
{\bf MAF} & $> 0.01$ &  $> 0.01$ & $> 0.01$ \\
&&&\\
\hline
&&&\\
{\bf ID Call Rate} & $ > 0.95 $ & $> 0.95$ & $ > 0.95$ \\
&&&\\
\hline
&&&\\
{\bf Missing Data} & Retained &  Retained & Retained \\
&&&\\
\hline
\hline
&&&\\
{\bf LD window} & 50 &  50 & 50\\
&&&\\
\hline
&&&\\
{\bf Shift} & 5 &  5 & 5\\
&&&\\
\hline
&&&\\
{\bf $r^2$} & $< 0.2$ &  $< 0.2$ & $< 0.2$ \\
&&&\\
\hline
\end{tabular}
\caption*{{\bf Table S1}: Parameters of the reduced dataset used for analysis}
\label{REDDAT}
\end{center}
\end{table*}

\bigskip
\bigskip

\begin{figure}[!ht]
\begin{center}
\includegraphics[width=8cm]{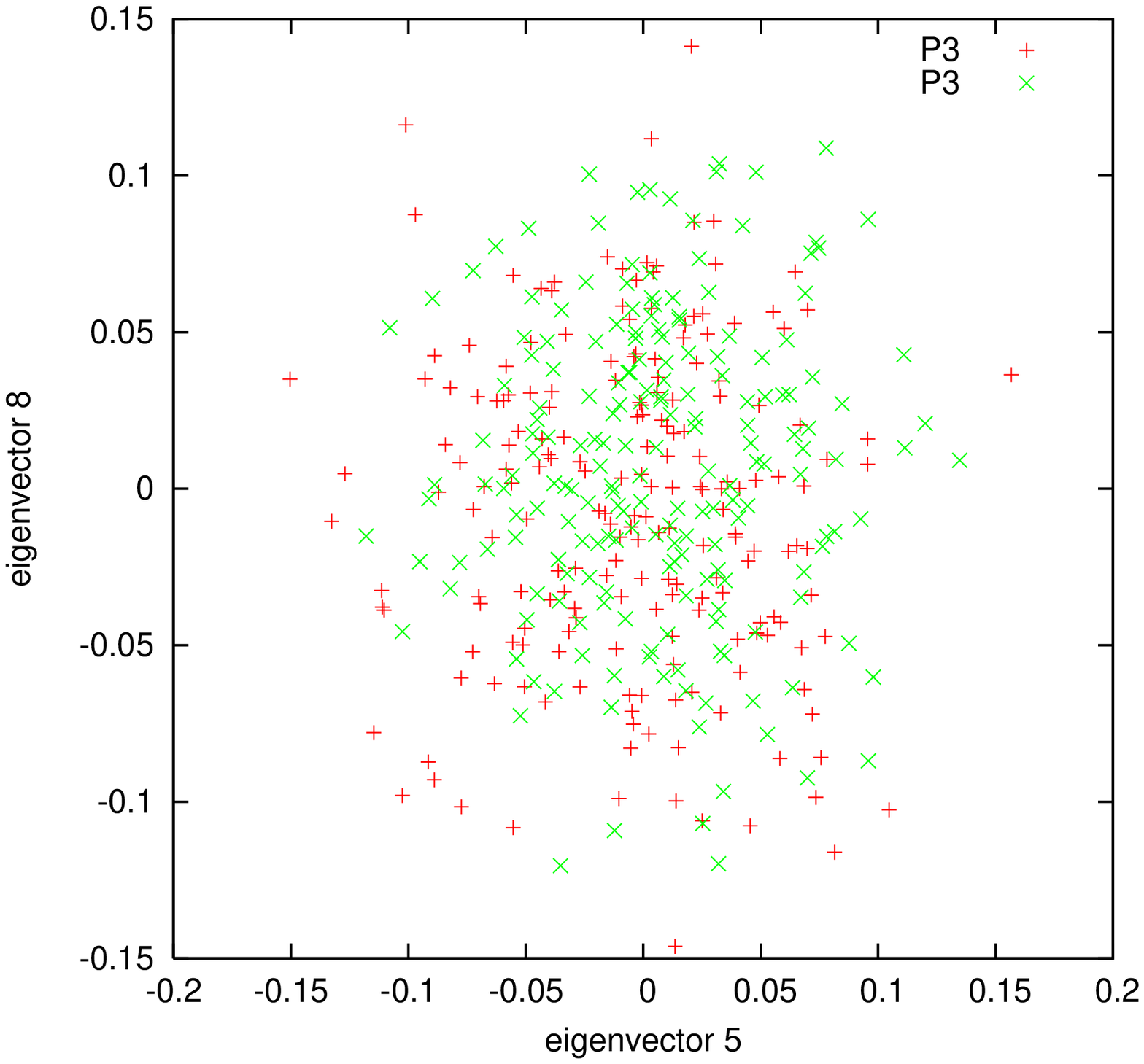}
 \caption*{{\bf Figure S1:} Intra- population projection of the P3 population (5739 SNPs, $p = 0.022$), along the two most significant axes
 . It is clear that despite the nominally significant $p$-value, the two sub-populations fail to separate along these axes.}
\label{P3P3ALL}
\end{center}
\end{figure}

\begin{table*}[!ht]
\begin{center}
\begin{tabular}{|c|c|c|c|c|c|c|c|c|}
\hline
&&&&&&&&\\
 $P_R$ &  $P_C$& $M_R$ & $M_C$ &  $N$ &  $F_{ST}(crit)$ &  $\hat{F}_{ST}$ &  SE &  Pval \\
&&&&&&&&\\
\hline\hline\hline
&&&&&&&&\\
{\bf P1} &{\bf P2} & 200 & 200 & 50 & 0.007 & $ < 0.0001$ & $ <0.0001$  & 0.725 \\
&&&&&&&&\\
\hline
&&&&&&&&\\
{\bf P1} &{\bf P3} & 200 & 200 & 50 & 0.007 & $ < 0.0001$ & $< 0.0001$  & $4.23 \times 10 ^{-4}$ \\
&&&&&&&&\\
\hline
&&&&&&&&\\
{\bf P2} &{\bf P3} & 200 & 200 &  50 & 0.007 & $ < 0.0001$ & $< 0.0001$  & 0.002 \\
&&&&&&&&\\
\hline
\hline
\hline
&&&&&&&&\\
{\bf P1} & {\bf P2} & 200 & 200 & 500 & $0.002$ & $ < 0.0001$ & $ 0.0002$  & 0.410 \\
&&&&&&&&\\
\hline
&&&&&&&&\\
{\bf P1} &{\bf P3} & 200 & 200 & 500 & ${\bf 0.002}$ & $ {\bf 0.0032}$ & $ 0.0003$  & ${\bf 1.75 \times 10 ^{-27}}$ \\
&&&&&&&&\\
\hline
&&&&&&&&\\
{\bf P2} &{\bf P3} & 200 & 200 & 500 & ${\bf 0.002}$ & $ { \bf 0.0026}$ & $ 0.0003$  & ${ \bf 6.32 \times 10 ^{-19}}$ \\
&&&&&&&&\\
\hline
\hline
\hline
&&&&&&&&\\
{\bf P1} &{\bf P2} & 200 & 200 & 5739 & $0.0007$ & $ < 0.0001$ & $ < 0.0001$  & 0.103 \\
&&&&&&&&\\
\hline
&&&&&&&&\\
{\bf P1} &{\bf P3} & 200 & 200 & 5739 & ${\bf 0.0007}$ & $ {\bf 0.0032}$ & $ 0.0002$  & ${\bf 2.86  \times 10 ^{-154}}$\\
&&&&&&&&\\
\hline
&&&&&&&&\\
{\bf P2} &{\bf P3} & 200 & 200 & 5739 & ${\bf 0.0007}$ & ${\bf  0.0031}$ & $ 0.0002$  & ${\bf 1.48 \times 10 ^{-155}}$\\
&&&&&&&&\\
\hline
\end{tabular}
\end{center}
 \caption*{{\bf Table S2:} PCA results for inter-population tests. $P_R$ and $P_C$ are the reference and comparison datasets, $M_R$ and $M_C$ the
 respective sample sizes and $N$ the number of SNPs used. $F_{ST}(crit)$ is the value of $F_{ST}$ at which the phase transition is
 expected. $\hat{F}_{ST}$ is the estimate of the $F_{ST}$ and SE is its standard error. Pval is the ANOVA $p$-value. The 50 SNP and 500 SNP sets were
 a contiguous set starting from the 1000th data point along the chromosome. Note the sharp drop in $p$-value at the BBP transition when $\hat{F}_{ST}$
 exceeds $F_{ST}(crit)$.}
\label{PCAINTER}
\end{table*}

\bigskip
\bigskip

\begin{figure}[!ht]
\begin{center}
\includegraphics[width=5cm]{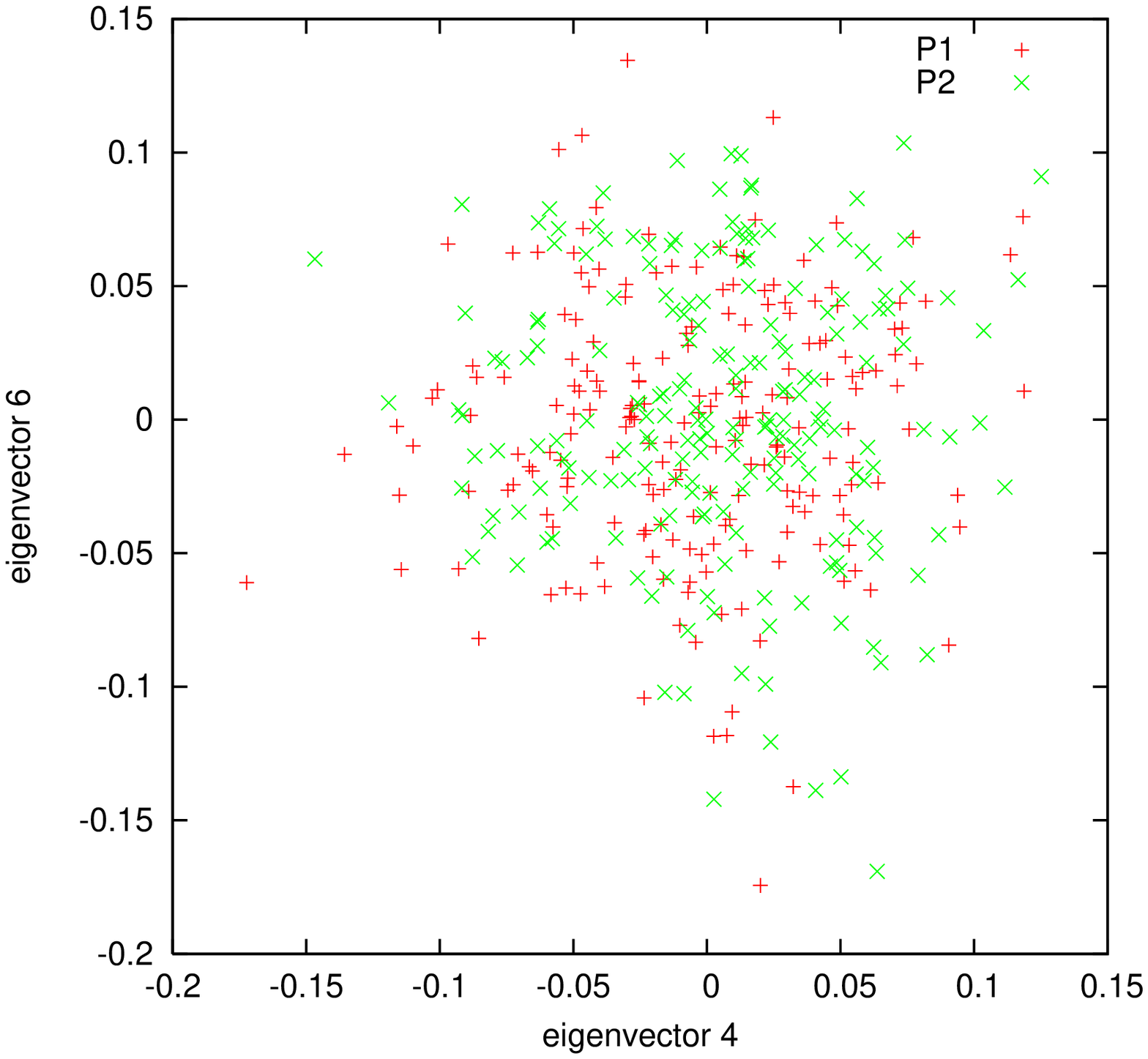}
\hfill
\includegraphics[width=5cm]{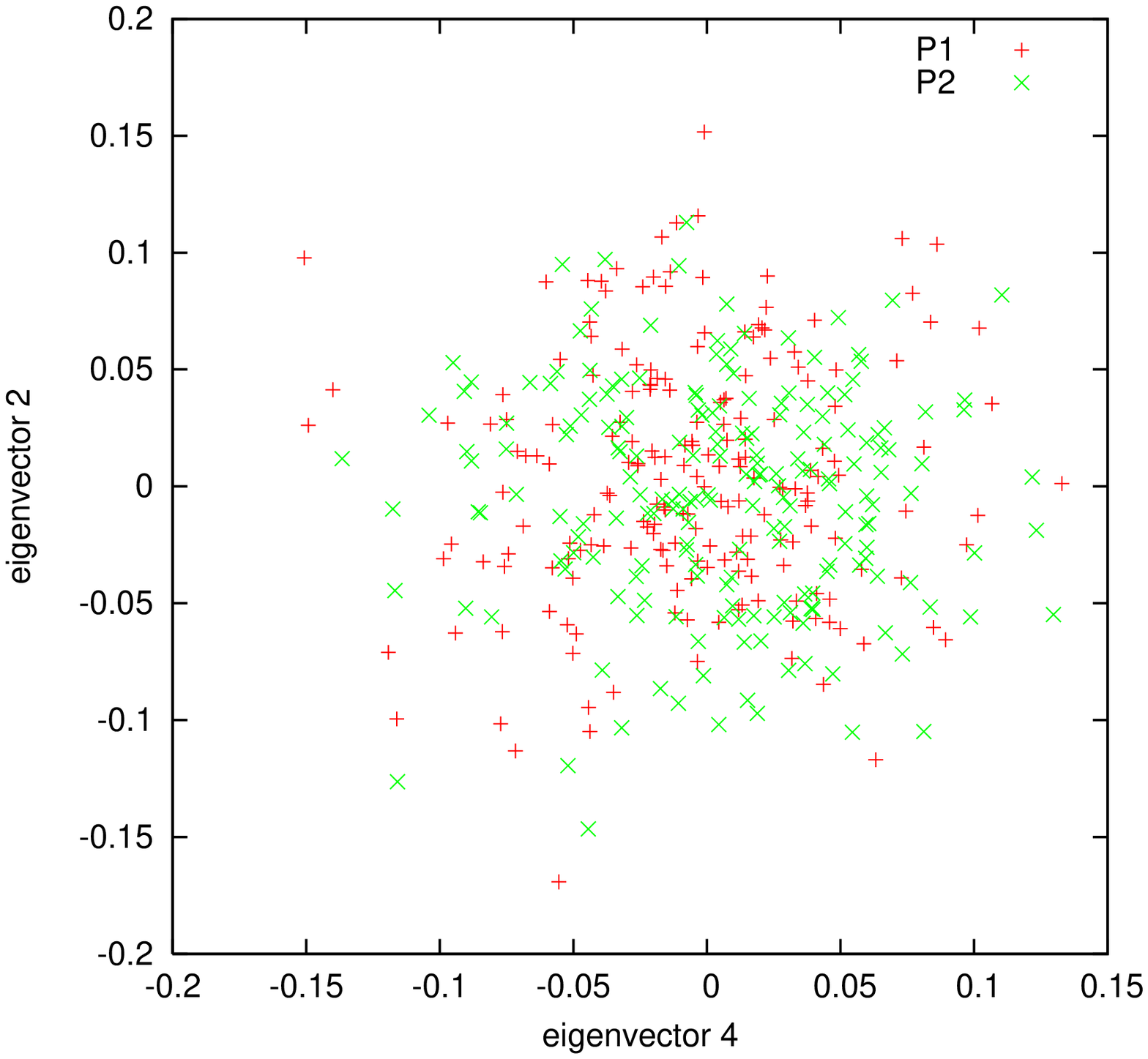}
\hfill
\includegraphics[width=5cm]{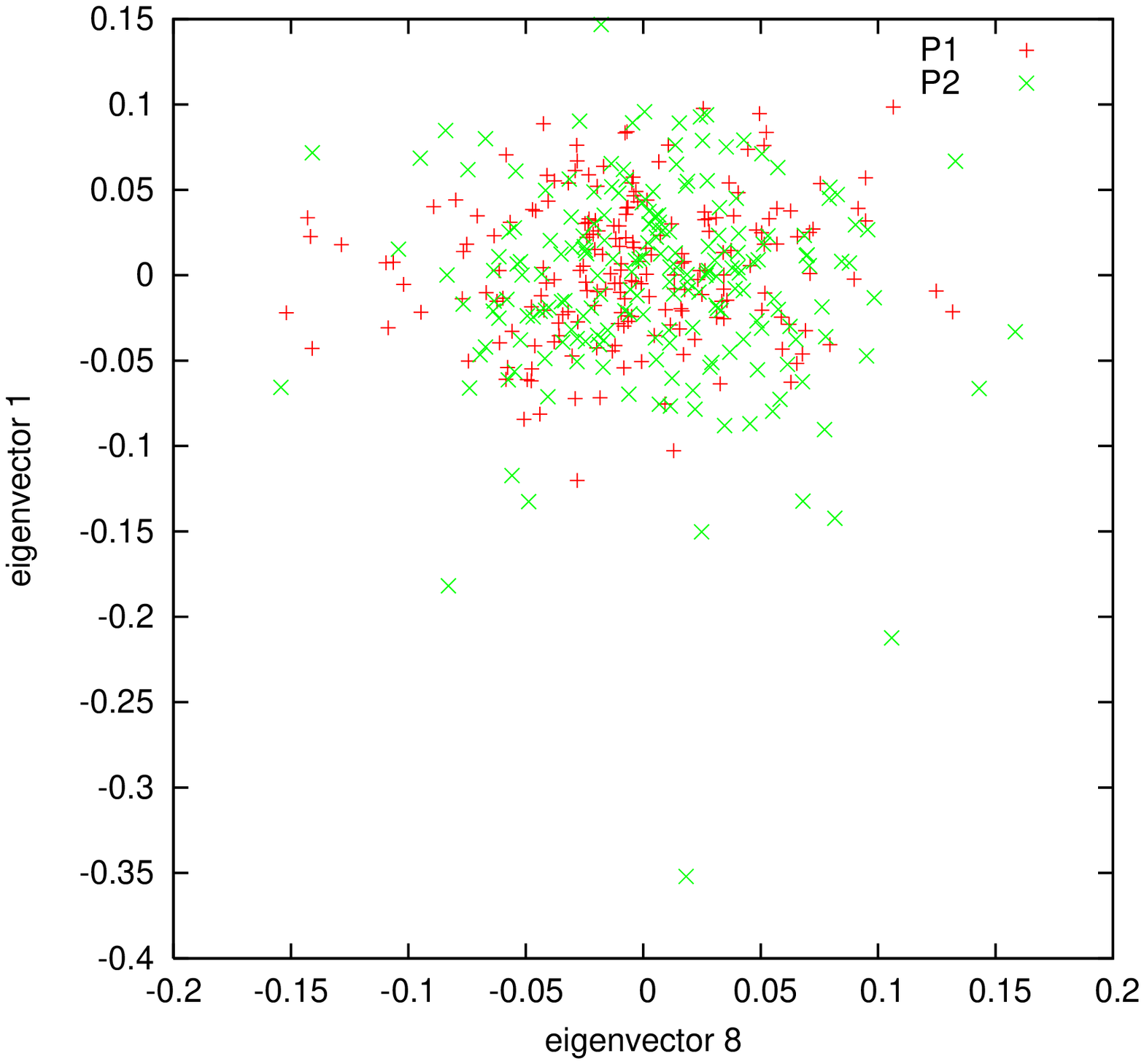}
\caption*{{\bf Figure S2:} Inter-population projection of the P1 and P2 population along the most significant axes for each value of
 $N$. $F_{ST}(crit)$ is never exceeded and the populations do not separate.}
\label{P1P2}
\end{center}
\end{figure}

\begin{figure}[!ht]
\begin{center}
\includegraphics[width=5cm]{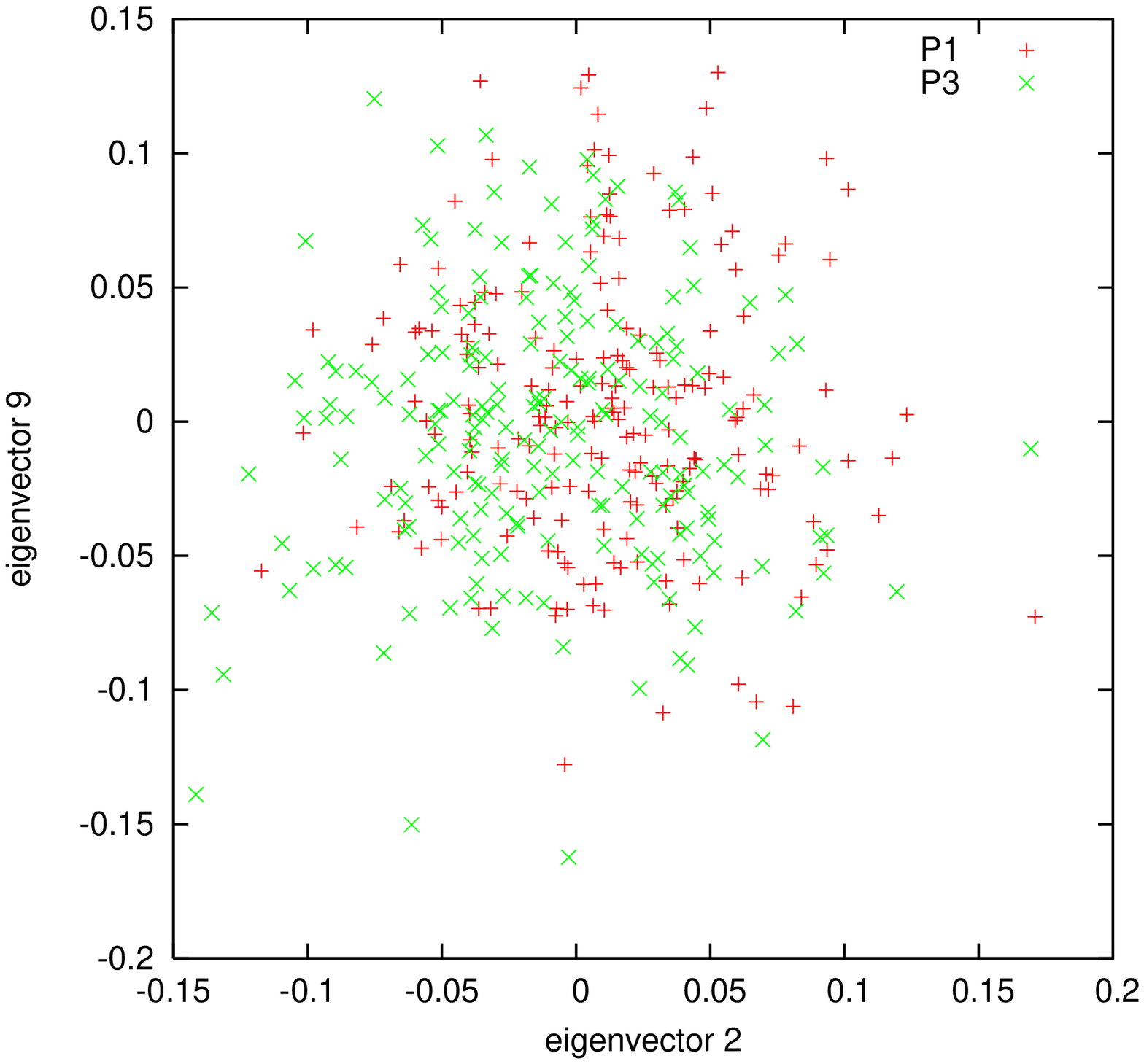}
\hfill
\includegraphics[width=5cm]{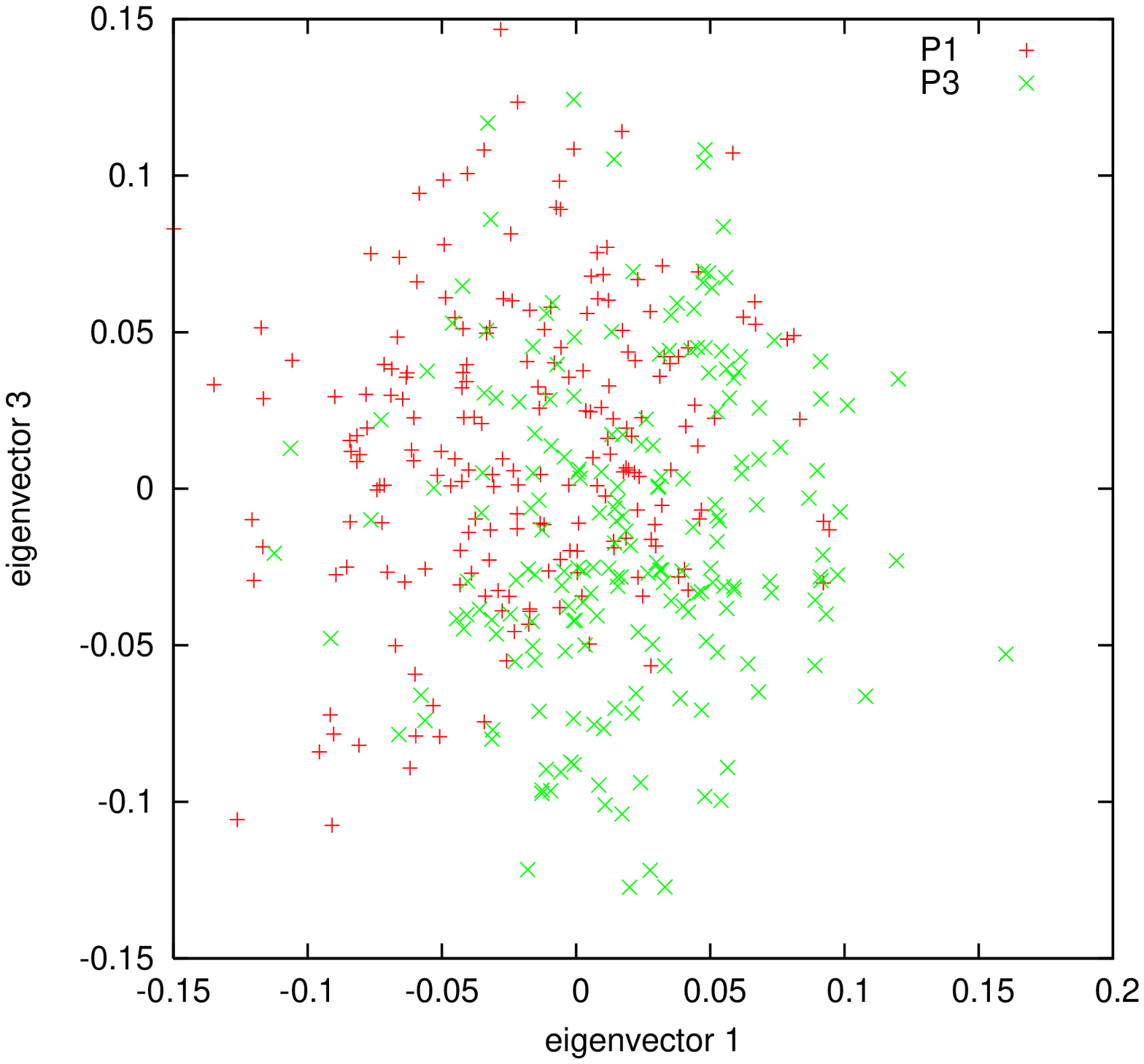}
\hfill
\includegraphics[width=5cm]{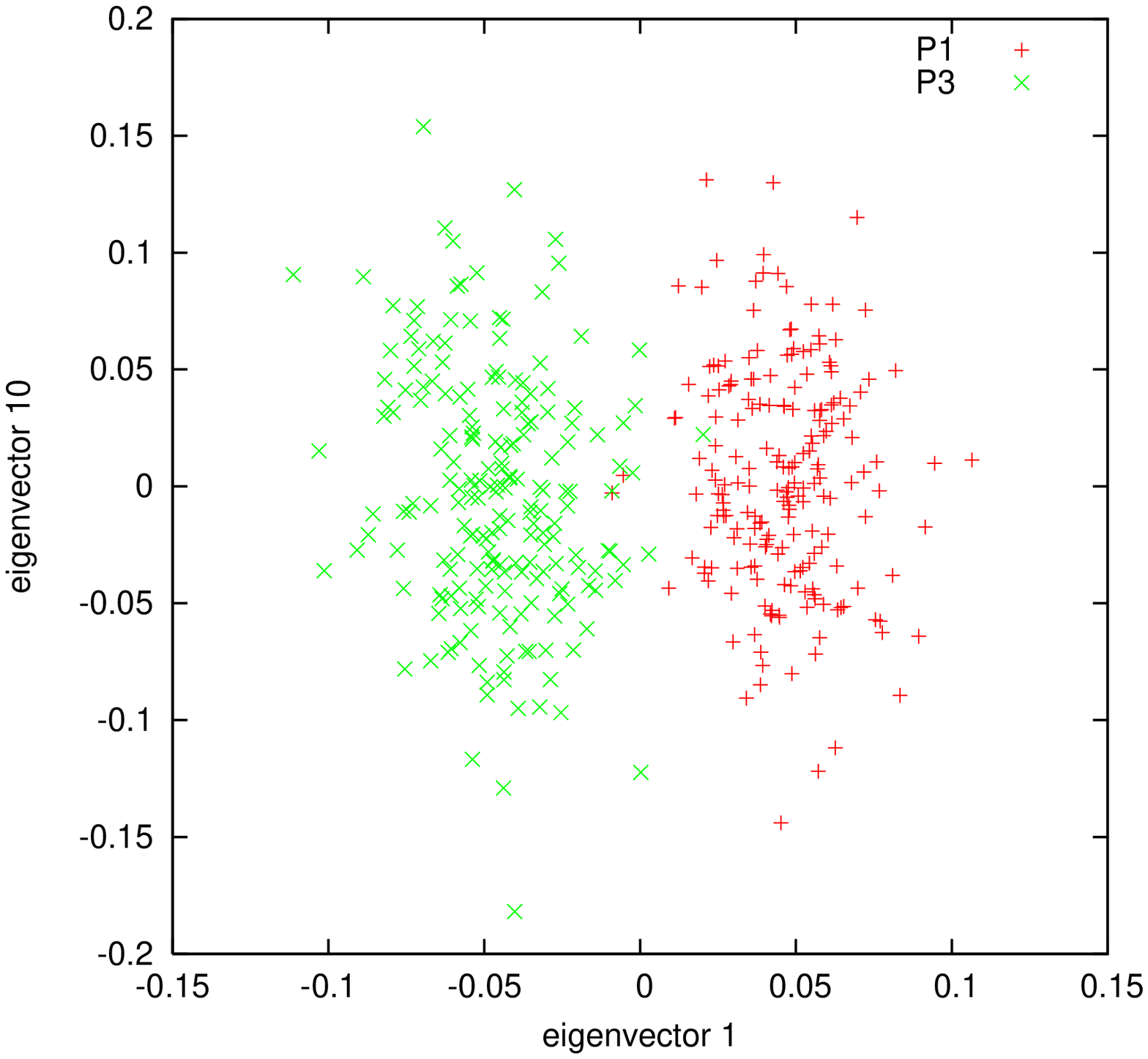}
 \caption*{{\bf Figure S3:} Inter-population projection of the P1 and P3 population along the most significant axes for each value of $N$. The
 populations separate as $F_{ST}(crit)$ is exceeded. }
\label{P1P3}
\end{center}
\end{figure}